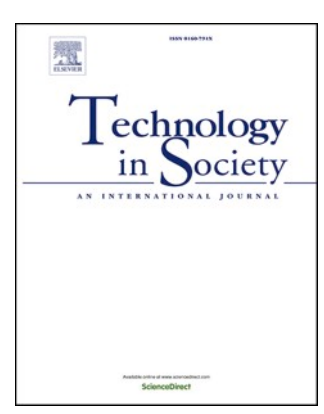

# Navigating the muddy waters of bias in artificial intelligence research: Understanding divergent meanings and conceptions

Mohammad Hossein Jarrahi [a], Amir Karami [b,*], Patrick Conway [a], Ali Memariani [c], Christoph Lutz [d,**]

[a] University of North Carolina at Chapel Hill, 302 Manning Hall, Chapel Hill, 27599, NC, USA
[b] School of Data Science and Analytics, Kennesaw State University, 1000 Chastain Road, Kennesaw, 30144, GA, USA
[c] Syngenta, 9 Davis Dr, Durham, 27703, NC, USA
[d] BI Norwegian Business School, Nydalsveien 37, Oslo, N-0484, Norway

ABSTRACT

As artificial intelligence (AI) pervades many decision-making domains, AI bias grows in importance. Although there is increasing awareness of the social and ethical consequences of biased AI, understanding bias from the perspective of those who develop these systems, such as the AI research community, is less clear. In this study, we employ topic modeling on 6520 articles to explore how the AI research community interprets the concept of bias. Our results show that the definition of bias is dispersed and complex within the community, often exhibiting even divergent conceptions (some even view and introduce bias as a tunable statistical parameter rather than an undesirable issue). The research community as a whole needs to engage more effectively with the concept of bias and establish a more cohesive understanding of it. We specifically argue that, although some sub-communities view bias as an issue that can be captured and mitigated through technical, computational, or statistical methods, it is not solely a technical problem. It instead involves contextual, social, and ethical factors that require broader sociotechnical perspectives and solutions.

## 1. Introduction

Artificial Intelligence (AI) and Machine Learning (ML) algorithms are constantly evolving and expanding in significance as they integrate into various applications, from human resource management to medical contexts (Jarrahi et al., 2023a, 2023b; Simon et al., 2023) . As AI spreads across domains, the impact of biased decision-making on people's lives becomes more severe (Kordzadeh & Ghasemaghaei, 2022).

Algorithmic bias has long been the concern of information researchers. Three decades ago, Batya Friedman and her colleagues developed methods to identify biases in computer systems and established the categories of "preexisting, technical, and emergent" bias (Friedman & Nissenbaum, 1996). Researchers have even argued that bias is an intrinsic aspect of AI systems, asserting that "without bias, all potential functions must be considered as hypotheses. Collectively, these functions predict that every possible future outcome has equal likelihood, making them unsuitable as a foundation for generalization or prediction" (Dietterich & Kong, 1995, p. 1). Since then, research has explored, defined, and advanced AI systems' potential to amplify existing biases in unpredictable, complicated ways (Ntoutsi et al., 2020). For example, a growing body of research has examined algorithmic fairness, accountability for outcomes, and bias detection processes (Mittelstadt et al., 2016).

Although conceptualizing and detecting bias in AI is important (Barocas & Hardt, 2023), there is still insufficient understanding of how the AI research community views and characterizes bias. The issue of bias gained substantial attention after 2010, with a growing body of research addressing algorithmic fairness, data bias, and ethics in AI (Barocas & Hardt, 2023; Kordzadeh & Ghasemaghaei, 2022; Mehrabi et al., 2022). Additionally, the availability of large datasets and regulatory frameworks post-2000 has made it possible to study and partly mitigate bias in contemporary AI technologies (Binns, 2018). While there have been several review articles and reference works on algorithmic bias, ML bias as well as the related issue of fairness in recent years (e.g., Barocas & Hardt, 2023; Ferrara, 2023; Hall & Ellis, 2023; Kordzadeh & Ghasemaghaei, 2022; Mehrabi et al., 2022; Pagano et al., 2023), these contributions focus either on specific forms of bias (e.g., Hall & Ellis, 2023 on gender bias), a specific AI application type or domain (e.g., Xue et al., 2023 on chatbots), or they are not systematic or comprehensive (Barocas & Hardt, 2023; Ferrara, 2023; Mehrabi et al., 2022). None of the existing reviews use computational methods to survey the complex literature on AI bias comprehensively. For example, while valuable in their insights, Kordzadeh and Ghasemaghaei (2022) analyzed only 56 articles, and Pagano et al. (2023) even fewer, with 45, limiting their coverage of the broader AI research field.

* Corresponding author.
** Corresponding author.
*E-mail addresses:* jarrahi@unc.edu (M.H. Jarrahi), akarami@kennesaw.edu (A. Karami), plconway@email.unc.edu (P. Conway), ali.memariani@syngenta.com (A. Memariani), christoph.lutz@bi.no (C. Lutz).

Across the existing review articles (and the AI bias literature more generally), a process focus on the ML pipeline is common, where bias can emerge because of the data, the model/algorithm, and the user interaction with the system, allowing for overlaps and feedback loops between the phases (Barocas & Hardt, 2023; Mehrabi et al., 2022). However, antecedents of bias tend to differ across phases. For example, some reasons for biased data include inconsistent sampling (Ferrara, 2023, sampling bias), faulty measurement (Mehrabi et al., 2022, measurement bias), and errors in data cleaning. When it comes to the model/algorithm, the specific ML approach can lead to biases, for example, when the model weights or parameter choice systematically advantage a certain group. Finally, bias in user interaction can occur once the AI system is deployed, for example, in the form of "generative bias" (Ferrara, 2023), where a model outputs content that is unbalanced or skewed (e.g., Gmyrek et al., 2025) or when users strategically manipulate a system to expose issues and biases (e.g., Microsoft's Twitter chatbot Tay, which had to be shut down after 24 hours because users managed to elicit highly offensive content in the form of sexist and racist tweets, see Wolf et al., 2017). Similar to the antecedents of AI bias, the consequences are complex and partly domain-dependent. Barocas and Hardt (2023) discuss many examples of biased AI systems and their downstream consequences in the form of discrimination, the potential of perpetuating stereotypes (e.g., relating to gender roles), inequalities, and tangible individual harms. Finally, some existing synthesis attempts also describe mitigation strategies to address biases, including pre-processing steps, model guidance, post-processing decisions (Ferrara, 2023), and emerging AI safety procedures (e.g., red teaming in the case of generative AI; see Feffer et al., 2024).

The AI research community serves as a key stakeholder in the AI ecosystem, and its understanding and treatment of bias is therefore highly consequential in how AI systems are developed and implemented. Computational research advancing deep neural networks, natural language processing, and information retrieval directly affects AI developers, engineers, and those implementing and using these systems (Simon et al., 2023). As these AI systems are rolled out in real-life and high-stakes settings, scholars pay increasing attention to the notion that computational and technical researchers are themselves "ethical agents" (Griffin et al., 2023), with an urgent need to understand how the AI research community approaches bias in AI. For example, Olteanu and colleagues (2019) argue that distinguishing between the normative connotations and the statistical understanding of bias presents a considerable challenge.

Consequently, a systematic and comprehensive examination of AI research that unpacks perspectives on bias offers insights into compatible or distinct understandings, encourages cross-community dialogue, and helps bridge conceptual gaps between AI researchers, social and ethical researchers, and other stakeholders. In this study, we employ computational literature review (CLR) as a structured and systematic approach to conducting systematic literature reviews, incorporating the mining of large text datasets within a specific knowledge domain (Antons et al., 2023). We rely on topic modeling and analyze AI articles that discuss bias, sourced from relevant journals and conferences. Topic modeling helps uncover the underlying structure of the data by clustering related terms into coherent topics (Vayansky & Kumar, 2020). This approach provides insights into recurring themes, which would be difficult to do manually in large datasets. Topic modeling also allows to focus on the most prominent topics, ensuring that the analysis remains comprehensive and interpretable. Our findings show 18 topics across four major AI bias themes, namely algorithmic bias (five topics), application-specific bias (six topics), data bias (four topics), and social and ethical bias (three topics). Cross-cutting through these themes and topics, two primary conceptions of bias appear: technical and sociotechnical. Technical bias emphasizes the algorithmic performance of AI systems, while sociotechnical bias addresses the social and ethical consequences of algorithm design.

## 2. Methods

Following Tranfield et al. (2003) and Collins et al. (2021), we developed and executed the CLR along these steps: (1) creation of a substantial corpus of scholarly literature related to bias in AI and ML (Section 2.1), (2) employment of advanced topic modeling methodologies to group literature into relevant topics (Section 2.2), and (3) analysis of the topics and reporting of the findings (Sections 2.3 and 3).

### 2.1. Data collection and preprocessing

Conference papers are highly esteemed in Computer Science (CS) and are often considered as important as journal publications. They undergo a rigorous peer-review process, with three to five reviewers, leading to high rejection rates and strong research impact. Unlike in other disciplines, where conferences typically feature less scrutinized abstracts, CS conferences publish full papers that attract significant attention and citations. Some CS conference papers are later expanded and published in journals for additional exposure (Freyne et al., 2010). Therefore, we considered both conference papers and journal articles in this study. Following the PRISMA process in Fig. 1 (Page et al., 2021), we targeted abstracts, publication year, title, and DOI from multiple peer-reviewed digital resources indexed in ACM, EBSCO, IEEE, Springer, Web of Science (WOS), and the proceedings websites of top computer science conferences, including AAAI, ACL, CVPR, ECCV, ICML, ICCV, and NeurIPS (Freyne et al., 2010). There is a notable overlap between the databases used in this study and those not explicitly included, such as PubMed. For example, while this study primarily focused on CS publications, our databases also cover a substantial portion of medical and health-related research, creating a significant intersection with PubMed and key medical conferences. Although we did not directly collect data from medical and health conferences such as MICCAI, their proceedings are frequently indexed across multiple platforms, including WOS, IEEE, ACM, and Springer. This extensive cross-indexing ensures that our dataset remains comprehensive and includes research contributions from both computer science and biomedical fields, even without directly integrating PubMed.

This process did not fully follow the chronological order of the PRISMA framework. However, it conceptually covered its core stages (identification, screening, eligibility, and inclusion) by integrating screening into the topic analysis phase. This ensured that irrelevant topics and their associated papers were excluded, keeping the final dataset aligned with the study's scope and objectives.

#### 2.1.1. Query

We searched for the term "bias" in the title, abstract, or keywords of documents across multiple databases, identifying 16,464 papers. These included 5364 from WOS, 5504 from ACM, 2699 from IEEE, 2178 from Springer, 169 from ECCV, 127 from NeurIPS, 110 from AAAI, 69 from ICCV, 68 from ICML, 66 from CVPR, 63 from ACL, and 47 from EBSCO.

#### 2.1.2. Identification

In the identification stage, we focused on English-written papers published after 2000 to account for recent developments in this area. Including only AI papers published after 2000 was essential because this period marks significant advancements in ML, deep learning, and neural networks, which are key to understanding bias in modern AI systems. We used the title and DOI to remove duplicates because of overlaps between these resources. At this stage, there were 14,355 papers.

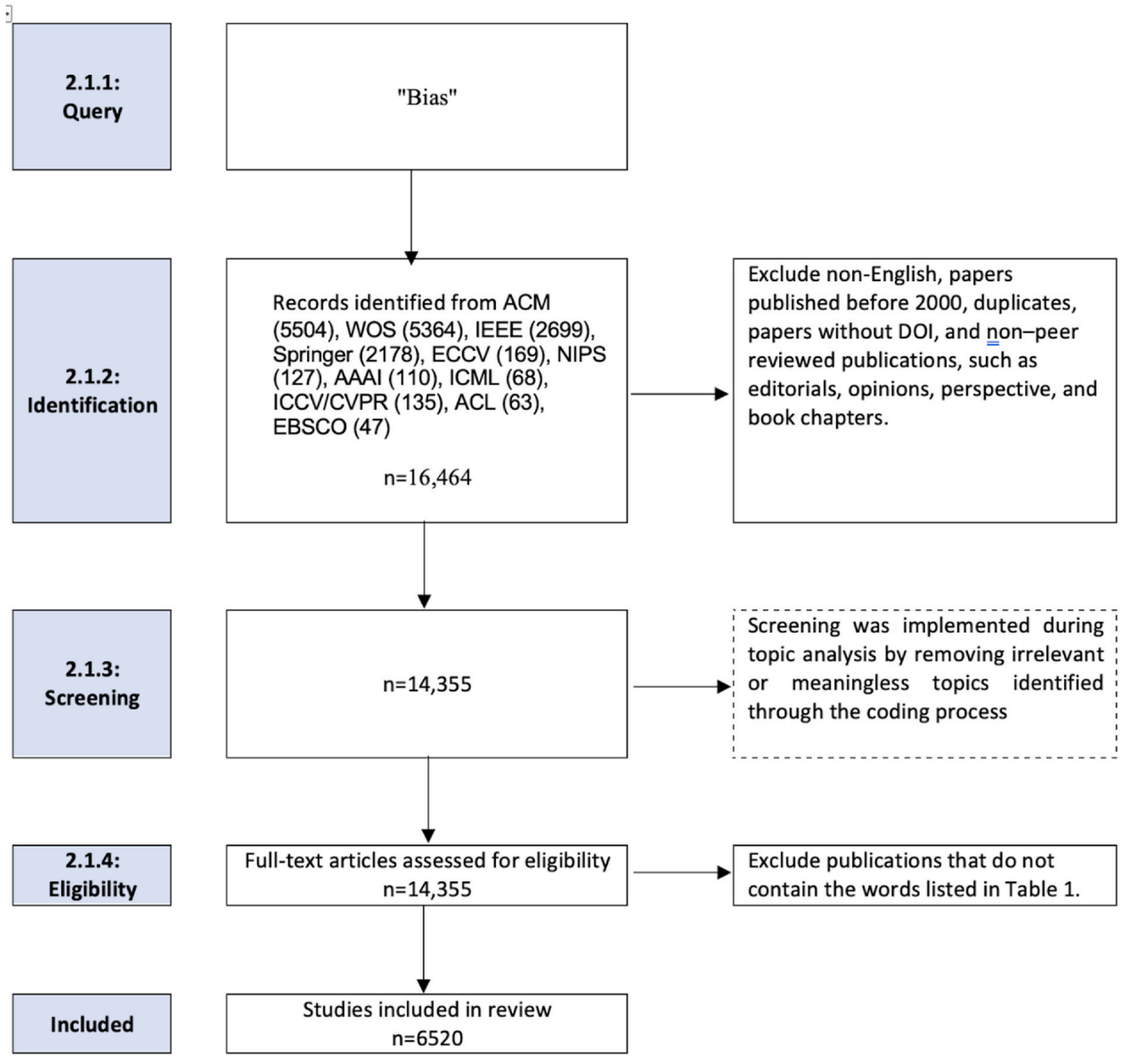

**Fig. 1.** PRISMA flow chart.

### 2.1.3. *Screening*

In this study, the screening stage was integrated into the topic analysis (Section 2.3) rather than implemented as a distinct chronological phase in the PRISMA workflow. During topic modeling, a coding process was applied to evaluate the semantic coherence and relevance of each discovered topic. Topics that were unrelated to the study's research questions or deemed too abstract, noisy, or off scope were systematically excluded. Because the removal of these irrelevant topics also eliminated the papers primarily associated with them, this process effectively functioned as both a topic-level and paper-level screening mechanism. While this procedure did not strictly follow the PRISMA framework in its traditional sequential form, it conceptually aligned with and fulfilled the intent of the PRISMA stages of identification, screening, eligibility, and inclusion by ensuring that only relevant and meaningful topics and papers were retained for analysis. This integrated approach allowed for a refinement of the dataset and ensured that the retained content accurately reflected the conceptual scope and objectives of the study.

### 2.1.4. *Eligibility*

In the eligibility stage, to align with the goal of identifying eligible studies, we developed a targeted search strategy to retrieve relevant papers containing "Bias" from the identification stage. We adopted three steps proposed by Webster and Watson (2022) to identify studies directly related to the research (Snyder, 2019). In the first step, with the help of the expert authors, we started with core AI and ML concepts, such as "machine learning" and "artificial intelligence." The focus at this step was on identifying broad concepts that could logically lead to specific algorithms or applications in subsequent stages. These core AI/ML concepts listed in Table 1 formed the basis of our initial search.

In the second step, we employed a backward citation analysis on a sample of 35 representative publications (five for each of the seven terms) obtained from Step 1 (i.e., studies retrieved using the core AI/ML keywords combined with "bias"). These initial papers served as anchor studies for identifying how bias was discussed within foundational and frequently used algorithms and methods. For each selected paper, we reviewed its reference list to trace the earlier works that the authors themselves had cited when discussing or applying AI/ML techniques.

**Table 1**
Search keywords.

| Bias + | | |
|---|---|---|
| Core AI/ML Concepts | Algorithm and Model Names | Applications and Methods |
| machine learning | artificial neural network | voice recognition |
| artificial intelligence | convnets | sound recognition |
| deep learning | convolutional neural networks | image recognition |
| computer vision | supervised learning | pattern recognition |
| natural language processing | unsupervised learning | text mining |
| NLP robotics | reinforcement learning | data mining |
| | semi-supervised learning | information retrieval |
| | bandit problem | sequence modeling |
| | classification model | transformer networks |
| | Clustering | denoising diffusion models |
| | multiagent systems | |
| | generative adversarial network | |

This backward tracing process allowed us to detect recurrent algorithmic and model-related terminology used in the context of bias or fairness. Through this iterative review, we noted that certain algorithms, such as artificial neural networks, reinforcement learning, supervised learning, and clustering models, appeared repeatedly in prior references discussing potential sources or mitigation strategies of bias. For instance, several anchor papers on “machine learning bias” cited foundational works on artificial neural networks that later informed the inclusion of this keyword in our list. This process both expanded and validated the keyword set by grounding it in existing literature networks rather than arbitrary inclusion. As a result, our keyword taxonomy grew to encompass a robust and evidence-based set of algorithmic and model descriptors, bridging the gap between general AI concepts and their technical instantiations in bias research.

The third step included a forward approach to explore a sample of 70 (two citations for each of the 35 anchor papers) cited in the literature identified in step 2, which expanded the list to include applications and methodologies, such as “voice recognition” and “data mining.” These keywords were used in conjunction with the term “bias” to account for all representations of bias, such as algorithmic bias and automation bias (Table 1). For example, using the query “machine learning” identified in the first step, we searched within the 14,355 papers and found a study on bias in classification (Tao et al., 2022). We then examined the citations within that paper and identified works related to “generative adversarial networks” algorithms (Fiore et al., 2019). Following the third step, we further discovered papers focusing on “data mining” methods (Bhattacharyya et al., 2011). Once this process was finalized, we stopped identifying new bias-related concepts in the literature, indicating that our list of keywords is complete and saturated. In summary, our three-step process produced seven core AI/ML concepts, twelve algorithms and models, and ten applications and methods, which collectively served as the keyword foundation for the eligibility stage (Table 1). This process resulted in a dataset of 6520 entries, including 3469 (53.2 %) journal articles and 3051 (46.8 %) conference proceedings articles.

### 2.2. Topic discovery

We employed topic modeling to identify topics frequently present in the 6520 articles. Topic modeling, a widely used method in text mining, has been used for various applications, such as computational literature mining for uncovering topics in fields such as management (Hannigan et al., 2019). Topic modeling is a more suitable method for our analysis of bias compared to traditional content analysis or grounded theory because 1) it can effectively manage large volumes of articles, 2) it eliminates the need for predefined dictionaries or interpretive rules, 3) it can uncover themes that might be challenging for human coders to explore and identify, and 4) it allows for grouping articles with similar perspectives on a concept such as bias (Vayansky & Kumar, 2020). While a citation network analysis is often employed in bibliometric studies and allows for meaningful insights (Waltman & van Eck, 2012), topic modeling ensures a more inclusive overview of the literature. It also avoids the issue of preferential attachment, where early or highly cited studies dominate the network regardless of their quality or novelty, potentially skewing the representation of the research landscape. From a feasibility standpoint, analyzing abstracts with topic modeling is more straightforward and scalable than constructing and processing large citation networks that rely on the availability and completeness of the references.

We applied the topic modeling approach to the abstract text. This approach is more efficient than analyzing the full text as it provides more comparable information (Cao et al., 2022). In addition, no substantial differences were found between topics generated from abstracts and full-text data based on topic quality using topic coherence analysis and human interpretation (Syed & Spruit, 2017), where extracting topics from abstracts is a more efficient approach than full-text analysis, as it can convey similar information using significantly fewer words (Cao et al., 2022).

Among various topic models, we employed Latent Dirichlet Allocation (LDA) to extract themes and identify the most pertinent articles for each (Blei et al., 2003). We selected LDA for multiple reasons. First, LDA has served as a reliable and established method to analyze large text data for discerning popular content, recognizing themes, and performing text categorization (Lu et al., 2011). LDA has been employed across different domains, such as politics/policy (Hagen, 2018), medical (Liu et al., 2022), business (Ao et al., 2023), information science (Miyata et al., 2020), management (Hannigan et al., 2019), and social media (Karami, Lundy, et al., 2020). This widespread adoption and robustness in diverse applications underscore LDA’s potential for topic identification in literature review studies.

Second, while more recent techniques like BERT-based approaches (Egger & Yu, 2022) offer certain advantages (e.g., contextual embeddings), LDA aligns well with the specific characteristics of our dataset and research objectives. In this case, LDA’s probabilistic framework, which does not rely on additional metadata or pre-trained language models, is appropriate for the scope and structure of our analysis (Egger & Yu, 2022). Over the last ten years, neural network models have become prominent in improving data analysis methods (Zhang et al., 2016). However, these models often require significant resources and time (Mohammadi et al., 2018) and need extensive training data, making them prone to overfitting (Erhan et al., 2010). Furthermore, deep learning techniques have not always surpassed traditional methods when dealing with small or medium-sized datasets (Mohammadi et al., 2018). LDA outperforms other topic models in certain datasets, demonstrating higher stability compared to other models (Altarturi et al., 2023). We compared the performance of LDA and BERTopic using topic coherence (TC), which quantifies the interpretability of a topic, and topic diversity (TD), which measures the percentage of unique words among the top words across all topics (Dieng et al., 2020). This evaluation was conducted on the 6520 papers retrieved during the identification step of our review process. The optimal topic coherence (TC) values for BERTopic and LDA were 0.02 and 0.36, respectively, while the topic diversity (TD) values were 0.72 for BERTopic and 0.95 for LDA. Since higher TC and TD values indicate better performance, this robustness makes LDA a reliable choice for extracting consistent topics, especially when working with diverse datasets.

Third, LDA is widely available and supported across multiple platforms, including Java,[1] Python,[2] and R.[3] This accessibility allowed for easy implementation and replication of the analysis across different technical environments. In contrast, newer methods like BERTopic are not universally supported and may require more advanced resources and technical expertise. Using LDA, we ensured that our methodology could be reused by a broad range of researchers without specialized infrastructure, making our findings more accessible and easier to reproduce. Using more advanced techniques in future research would be valuable, particularly when the inclusion of metadata or contextual information is necessary. Still, LDA's simplicity, flexibility, and comparative performance made it the right choice for this research.

[1] https://mimno.github.io/Mallet/topics.html.
[2] https://radimrehurek.com/gensim/.
[3] https://cran.r-project.org/web/packages/topicmodels/index.html.

In LDA, every document is a blend of topics, with each topic constituting a collection of semantically related words that denote high-level concepts. Applying LDA to a corpus yields two primary outcomes: the topics themselves and their corresponding documents (Karami, 2015). LDA can also depict the weight assigned to each topic within a document, illustrating the degree of connection between the topic and the document. The cumulative topic weights indicate how frequently each topic appears across the documents (Karami, White, et al., 2020). To identify the optimal set of topics, we employed a method developed by Deveaud et al. (2014) that accentuates the disparities in information between pairs of topics. This method ensured minimal overlap and the emergence of distinct topics. The procedure incorporated Latent Concept Modeling (LCM) to delineate latent concepts and LDA to highlight the most relevant topics derived from feedback documents (Deveaud et al., 2014). LCM helps to determine the optimum number of topics by assessing the coherence and relevance of the discovered topics to evaluate how well-related terms cluster together within topics. This coherence metric is calculated for different numbers of topics, and the model then selects the number of topics that maximizes this coherence score. In our case, LCM determined the optimum set to be 39 topics.

We used MALLET (MAchine Learning for LanguagE Toolkit) (McCallum, 2002), which is an open-source Java-based toolkit widely used for topic modeling. We applied LDA on abstracts with 39 topics, 4000 iterations, beta = 0.01, and alpha = 0.13 (five divided by the number of topics or 5/39). The number of iterations was set to 4000 to ensure full convergence and account for the random initialization in LDA, even though the log-likelihood had stabilized before reaching that point (Van Altena et al., 2016). To evaluate the robustness of LDA, we compared the log-likelihood for three sets of 4000 iterations and found that there was no significant difference between the mean and standard deviation of iterations. The random sampling process of LDA could lead to unstable topics that change after each running (Shao et al., 2016). To identify stable topics, we used cosine similarity to find common topics among the three sets of topics. We set the threshold at 0.5 and used the top 20 words in each topic to identify stable topics. A stable topic is the one that can be mapped to a topic in the three sets of topics (Shao et al., 2016). We also employed cosine similarity analysis to gauge the similarity among the 39 topics. This analysis revealed a mean similarity of 0.12 and a median of 0.10, signifying minimal overlap between the topics. This process identified 24 stable topics and excluded 15 unstable topics.

### 2.3. Topic analysis

To analyze the meaning of each topic, we developed a three-step human coding scheme by studying the top words in each topic and the most related full-text papers of each topic (Karami et al., 2024). First, for each topic, two authors independently answered the following yes/no questions: (Q1) "Does the topic present a meaningful and cohesive theme?" (Q2) "Is the topic related to the concept of bias?" These two questions determined whether a topic is meaningful and relevant. Cohesiveness (Q1) was assessed based on whether the keywords and the most representative papers reflected a clear, interpretable theme, while relevance (Q2) was determined by the topic's connection to bias-related concepts, such as bias in machine learning classifiers. This validation process led to the removal of six incohesive or irrelevant topics and ensured that the remaining topics (see Table 2 for the remaining 18 topics) were thematically coherent and aligned with the study's focus on bias in computational research. The agreements for meaningfulness and relevance were 91.7 % and 75 %, respectively. Second, two coders independently developed definitions and assigned labels for each topic. After this initial coding, both coders met to discuss their interpretations and resolve any differences. A third coder reviewed their definitions and decisions to confirm whether consensus had been reached. If all three coders agreed on a label for a topic, it was recorded as a consensus. Out of the 18 topics, consensus was reached on 88.9 %. Although the coders did not always use identical wording for topic labels, they consistently employed semantically similar terms and concepts. For example, one coder labeled Topic 2 as "classification algorithms," while the other referred to it as "machine learning classifiers." Despite the lexical differences, both descriptions conveyed the same underlying concept. In such cases, the coders discussed the similarities in meaning, aligned their interpretations, and collaboratively developed a unified label that accurately captured the shared intent. This process ensured consistency and conceptual clarity across all topic labels.

To address the disagreements, the two coders conducted another round of coding, and a third coder resolved any remaining disagreements. Associated keywords and their corresponding papers assisted the coders in determining the meaningfulness, relevance, and underlying theme of this topic. The coders used thematic analysis for interpreting the conceptual meanings of topics (Braun & Clarke, 2006). The coders independently and inductively reviewed the top-20 most relevant papers for each of the 18 topics and consequently determined a definition for each topic. Coders met to discuss the definitions until they agreed on the definitions. There were agreements over most definitions. However, in some cases the wording differed but was semantically consistent, and in other cases the coders did not agree. In the latter situation, the coders kept discussing until they reached an agreement.

## 3. Results

We conducted a close analysis of the 18 topics in Table 2, identifying the key research areas and the varying ways bias is understood across these topics. We also included a table listing the most frequently occurring journals and conferences for each topic in Appendix I. This table highlights the top venues where research related to each topic is published, providing insights into key publication outlets and their relevance within the dataset.

The 18 topics span a broad spectrum of AI/ML-related areas, and each focuses on the roles and nature of bias within its respective research domain. These range from technical issues like domain adaptation and classification algorithms to social and ethical concerns such as fairness, discrimination, and societal implications in AI systems. This analysis helps reveal how bias manifests differently depending on the research context, taking stock of both technical and sociotechnical perspectives. Fig. 2 illustrates the weight of the topics ordered from the most popular to the least popular. Domain adaptation, AI ethical issues, and classifications emerged as the most prevalent (popular), while information retrieval, AI medical applications, and cybersecurity were the least covered.

We also analyzed how often each topic appeared as the primary topic across all papers. For each paper, the primary topic was identified as the one with the highest weight or the highest probability P(T|D), meaning it was the most strongly associated topic for that paper. For example, if a paper exhibited varying topic probabilities across the 18 topics, the topic with the greatest probability value was designated as its primary topic. In this case, Topic 7 would be identified as the primary topic for that paper. We repeated this process for all papers and counted how many times each topic was identified as the primary topic across the entire dataset. These counts, representing the frequency of primary topics, are shown in Fig. 3. While a paper may be related to multiple topics, only one can be designated as the primary topic. Fig. 3 highlights that the top

**Table 2**
Top words of topics.

| ID | List of Relevant Words in the Topic | Most representative article |
|---|---|---|
| T1 | learning domain datasets model adversarial training bias deep propose target proposed dataset data performance generative loss models recognition method problem | Yang et al. (2023). Adaptive bias-aware feature generation for generalized zero-shot learning. IEEE Transactions on Multimedia, 25, 280–290. |
| T2 | class data classification learning imbalanced classes imbalance datasets performance samples problem minority classifiers majority proposed methods machine method results training | Chawla et al. (2003). SMOTEBoost: Improving prediction of the minority class in boosting. In Knowledge Discovery in Databases: PKDD 2003: 7th European Conference on Principles and Practice of Knowledge Discovery in Databases, Cavtat-Dubrovnik, Croatia, September 22–26, 2003. Proceedings 7 (pp. 107–119). Springer Berlin Heidelberg. |
| T3 | user users recommendation information model items recommender systems based social item reviews biased data recommendations product collaborative content behavior feedback | Xu et al. (2019). NeuO: Exploiting the sentimental bias between ratings and reviews with neural networks. Neural Networks, 111, 77–88. |
| T4 | network neural model ann artificial algorithm weights parameters proposed learning performance training hidden results input based biases optimization error elm | Cicek & Ozturk (2021). Optimizing the artificial neural network parameters using a biased random key genetic algorithm for time series forecasting. Applied Soft Computing, 102, 107091. |
| T5 | bias fairness learning models machine biases data groups datasets fair algorithms discrimination attributes model biased sensitive algorithmic group race mitigate | Hu & Rangwala (2020). Metric-free individual fairness with cooperative contextual bandits. In 2020 IEEE International Conference on Data Mining (ICDM) (pp. 182–191). IEEE. |
| T6 | neural deep network networks convolutional training learning cnn accuracy architecture performance bias model cnns classification trained architectures loss layers models | Sakai, Y. (2020). Quantization for deep neural network training with 8-bit dynamic fixed point. In 2020 7th International Conference on Soft Computing & Machine Intelligence (ISCMI) (pp. 126–130). IEEE. |
| T7 | detection network dataset attacks detect machine attack anomaly security false learning traffic malware accuracy model event detecting based events intrusion | Moghanian et al. (2020). GOAMLP: Network intrusion detection with multilayer perceptron and grasshopper optimization algorithm. IEEE Access, 8, 215202–215213. |
| T8 | algorithm algorithms problem optimization problems search results solutions set paper proposed solution approach selection optimal evolutionary techniques process genetic design | Fu et al. (2022). Improving probability selection based weights for satisfiability problems. Knowledge-Based Systems, 245, 108572. |
| T9 | studies health risk clinical methods data study treatment bias results patient care machine prediction outcome outcomes assessment potential medical patients | Suri et al. (2022). Understanding the bias in machine learning systems for cardiovascular disease risk assessment: The first of its kind review. Computers in biology and medicine, 142, 105204. |
| T10 | information retrieval search ranking query documents relevance web document results relevant systems queries effectiveness evaluation based paper performance experiments collections | Lam-Adesina and Jones (2001). Applying summarization techniques for term selection in relevance feedback. In Proceedings of the 24th annual international ACM SIGIR conference on Research and development in information retrieval (pp. 1–9). |
| T11 | learning data training labels labeled supervised classification label methods model semi-supervised unlabeled bias samples propose annotation method framework labeling set | Wu et al. (2022)). Label Aggregation with Clustering for Biased Crowdsourced Labeling. In Proceedings of the 2022 14th International Conference on Machine Learning and Computing (pp. 165–169). |
| T12 | systems intelligence artificial human machine ethical decisions design biases learning decision-making challenges research bias technology paper potential algorithmic explanations social | Malek, M. A. (2022). Criminal courts' artificial intelligence: the way it reinforces bias and discrimination. AI and Ethics, 2(1), 233–245. |
| T13 | image segmentation noise images bias method proposed algorithm field model intensity results local function based clustering paper fuzzy brain methods | Juntu et al. (2005). Bias field correction for MRI images. In Computer Recognition Systems: Proceedings of the 4th International Conference on Computer Recognition Systems CORES'05 (pp. 543–551). Springer Berlin Heidelberg. |
| T14 | models model learning prediction machine performance data regression accuracy training bias random ensemble decision predictive predictions predict algorithms tree methods | Islam and Hassanzadeh Amin (2020). Prediction of probable backorder scenarios in the supply chain using Distributed Random Forest and Gradient Boosting Machine learning techniques. Journal of Big Data, 7(1), 65. |
| T15 | learning reinforcement policy agents reward action algorithm agent actions bias environment learn control task problem deep policies algorithms tasks performance | Swazinna et al. (2021). Overcoming model bias for robust offline deep reinforcement learning. Engineering Applications of Artificial Intelligence, 104, 104366. |
| T16 | language text bias natural processing social sentiment content speech nlp analysis information semantic corpus model political task approach research models | Fast et al. (2016). Shirtless and dangerous: Quantifying linguistic signals of gender bias in an online fiction writing community. In Proceedings of the International AAAI Conference on Web and Social Media (Vol. 10, No. 1, pp. 112–120). |
| T17 | clustering clusters data cluster algorithm similarity proposed method results based distance bias approach set measure k-means algorithms similar unsupervised propose | Gharghabi et al. (2018)). Matrix profile xii: Mpdist: a novel time series distance measure to allow data mining in more challenging scenarios. In 2018 IEEE International Conference on Data Mining (ICDM) (pp. 965–970). IEEE. |
| T18 | image images vision visual computer object objects detection recognition model dataset features color saliency attention bias shape scene information human | Tseng et al. (2009). Quantifying center bias of observers in free viewing of dynamic natural scenes. Journal of vision, 9(7), 4-4. |

five primary topics (i.e., Domain Adaptation, Bias Shift, Predictive Modeling, AI Ethical Issues, and Classification) were identified as the primary focus in 2485 papers.

### 3.1. General overview: overarching AI bias themes

For conceptual clarity and to narrow down the complexity, the 18 identified topics can be grouped into four AI bias themes. These themes help illustrate how bias has been conceptualized, problematized, and operationalized in AI research: (1) **Algorithmic Bias (Bias in Model & Learning Process)**, which focuses on the design, optimization, and training of AI systems, and typically frames bias as a statistical inefficiency or a technical parameter to be minimized or fine-tuned for better systems performance (Topics 2, 4, 14, 15, and 17); (2) **Application-Specific Bias**, which addresses the manifestation of bias within specific domains and application contexts such as healthcare, computer vision, and medical imaging, and focuses on the impact of contextual and application-specific factors in interpreting and managing bias (Topics 7, 9, 13, 10, 16, and 18); (3) **Data Bias (Issues in Dataset & Representation)**, which embodies concerns related to the quality, distribution, annotation, and representativeness of training data, and focuses on how biases introduced in data pipelines and labeling processes may adversely impact AI outcomes (Topics 1, 6, 8, and 11); and (4) **Social and Ethical Bias**, which mostly draws attention to issues of fairness, discrimination, accountability, and the broader ethical and social consequences of AI systems when adopted in real-world (Topics 3,

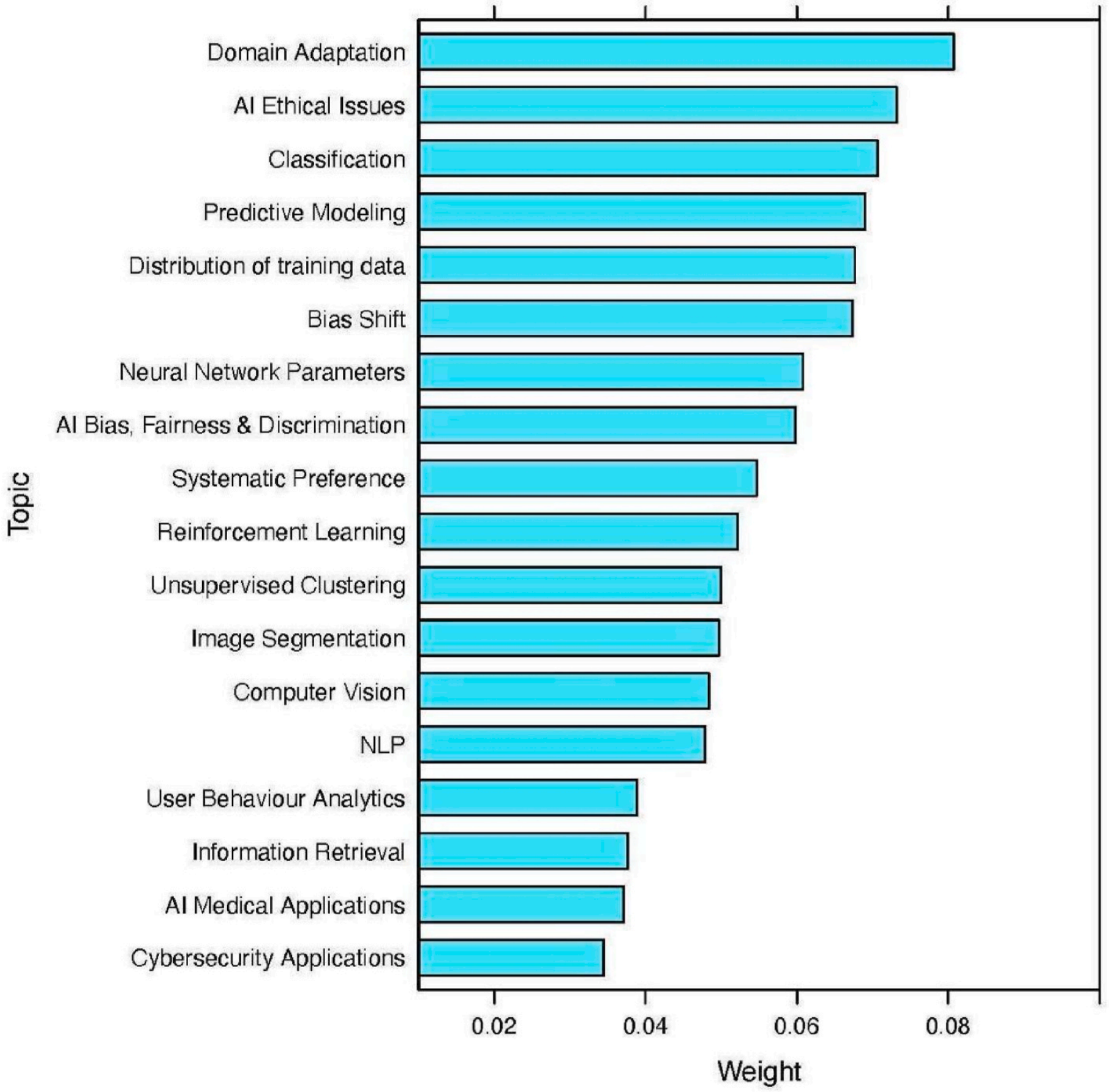

**Fig. 2.** Weight of topics.

5 and 12).

To further contextualize the topics, we draw on a widely used typology of bias that has been formulated in the literature on ML and algorithmic fairness (Barocas & Hardt, 2023; Mehrabi et al., 2022; Schwartz et al., 2022; Suresh & Guttag, 2021). This typology includes six major bias types (see Table 3), helping to systematically classify how bias emerges at different stages of the machine learning pipeline.

Each type reflects a distinct mechanism by which bias can arise, ranging from entrenched social inequalities in data (historical bias) to the assumptions embedded in model evaluation and real-world deployment (evaluation and deployment bias). We use these types as a point of comparison to better situate the 18 research topics identified in our analysis. Table 4 demonstrates the bias types most closely associated with the four themes from our analysis.

While there is substantial overlap between our themes and existing bias types in the literature, several points of divergence exist. First, certain topics, especially Topic 4 (Neural Network Parameters), Topic 14 (Predictive Modeling) and Topic 15 (Reinforcement Learning), represent technical artifacts in model training or reinforcement learning that are not adequately represented in pipeline-based classifications such as those summarized in Table 3 (Mehrabi et al., 2022; Suresh & Guttag, 2021). Second, some topics (e.g., T8 Systematic Preference and T17 Unsupervised Clustering) show that AI bias is not always conceptualized as a flaw but as a strategic design choice or modeling heuristic (in contrast to the dominant framing of bias as a negative factor to be diminished or eliminated). Third, the social and ethical topics (T3 User Behavior Analytics, T5 AI Bias, Fairness & Discrimination, and T12 Ethical Issues) expand the scope of bias beyond the procedural stages in what we call standard typologies[4] (see Table 3, the six widely adopted bias categories by Suresh and Guttag (2021) and subsequent frameworks by Mehrabi et al. (2022) and Schwartz et al. (2022)). This happens by highlighting broader sociotechnical concerns such as discrimination, user feedback loops, and institutional accountability. Finally, the domain-specific manifestations of bias (T9 AI Medical Applications and T18 Computer Vision) further illustrate bias as strongly shaped by contextual factors not fully accounted for in standard typologies. These discrepancies suggest that while existing typologies offer a helpful foundation, they require expansion to accommodate the evolving and multifaceted ways AI bias is understood across research domains.

In the following, we will use the four overarching themes as the main structuring device to discuss the results, with the themes serving as subsections and the 18 topics as sub-sub-sections nested within each theme. The first two themes (3.2 algorithmic bias and 3.3 application-specific bias) capture a more technical understanding, while the last two themes (3.4 data bias and 3.5 social and ethical bias) refer to a more sociotechnical understanding. We will come back to this distinction in the Discussion.

[4] These typologies are considered “standard” because they have achieved wide recognition and uptake across the AI fairness and bias literature. For instance, Suresh and Guttag (2021) has 875 Google Scholar citations as of 24 October 2025, Mehrabi et al. (2022) 7820 citations and Schwartz et al. (2022) 627 citations. Thus, they are among the most frequently cited and foundational works guiding subsequent classifications of bias types in AI.

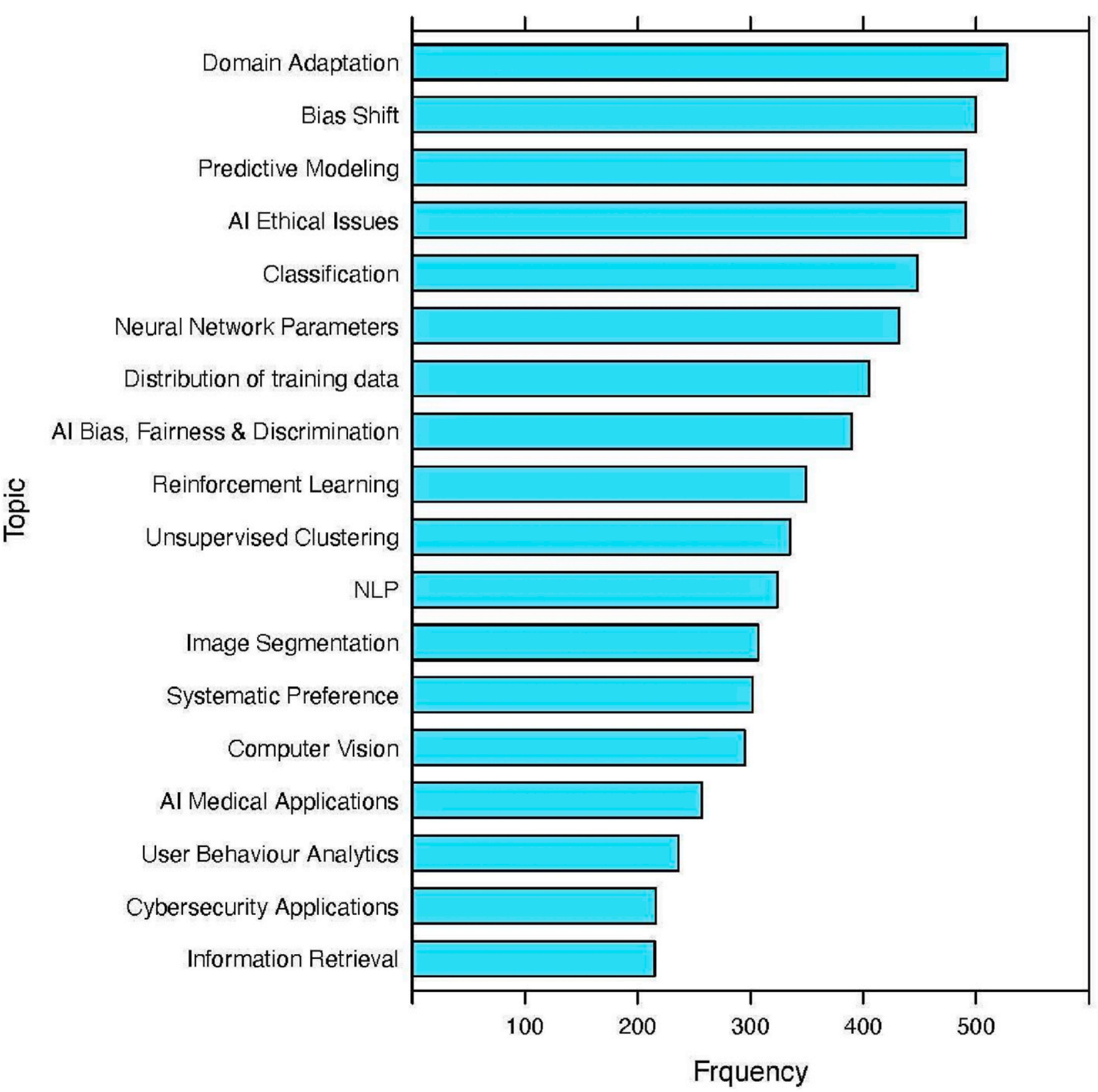

**Fig. 3.** Frequency of papers by primary topic.

**Table 3**
Types of AI bias in existing research.

| Bias Type | Definition | Key Sources |
|---|---|---|
| Historical Bias | Bias arising from societal inequities already embedded in real-world data, even if perfectly sampled. | – Suresh and Guttag (2021)<br>–Mehrabi et al. (2022)<br>–Schwartz et al. (2022) |
| Representation Bias | When some groups or characteristics are under- or overrepresented in the training data. | –Suresh and Guttag (2021)<br>–Mehrabi et al. (2022)<br>–Schwartz et al. (2022) |
| Measurement Bias | When features or labels used in modeling are imperfect proxies for real-world concepts of interest. | –Suresh and Guttag (2021)<br>–Mehrabi et al. (2022) |
| Aggregation Bias | When model aggregates data across heterogeneous groups with a "one-size-fits-all" mindset. | –Suresh and Guttag (2021)<br>–Mehrabi et al. (2022) |
| Evaluation Bias | When model evaluation or test data does not reflect the real-world settings and criteria. | –Suresh and Guttag (2021)<br>–Mehrabi et al. (2022)<br>–Schwartz et al. (2022) |
| Deployment Bias | When contingencies of the deployment context differ from the assumptions made during its design and training. | –Suresh and Guttag (2021)<br>–Mehrabi et al. (2022)<br>–Schwartz et al. (2022) |

### 3.2. *Algorithmic bias: topics 2, 4, 14, 15, 17*

The algorithmic bias theme consists of five topics (see Table 5). All these topics share a technical understanding of bias in concrete AI implementation, particularly the development and use of ML algorithms.

#### 3.2.1. *Topic 2: classification algorithms*

This technical topic is focused on imbalanced datasets and how they impact classification methods. Imbalanced datasets, or datasets that contain smaller numbers of minority classes, "pose hindrance to conventional learning methods" (Raghuwanshi & Shukla, 2020, p. 1) and tend to classify more accurately towards the majority class. The studies under this topic propose technical methods to identify and address this specific definition of bias. While balanced training datasets are important for robust classification, it is increasingly evident that real-world applications often deviate from this requirement (Kumar et al., 2019). The technical methods explored in this topic, such as oversampling the minority classes (Chawla et al., 2003), generating synthetic examples of majority and minority classes to create a balanced dataset (Guo & Viktor, 2004), and incorporating advanced bagging and boosting methods (Kumar et al., 2019), have demonstrated improved prediction accuracy on minority classes.

Bias discussed in this topic is viewed as a challenge during classification algorithm development. While the research falling within this topic is mostly of an algorithmic-technical nature, the consequences of imbalanced datasets can be far-reaching in real-world applications; such datasets can potentially reinforce or even exacerbate existing inequalities, for instance, by under-representing minority classes in healthcare, finance, and criminal justice.

**Table 4**
Comparison of our AI bias themes and bias types in the literature.

| Themes from our analysis | Topic Numbers and Titles | Corresponding bias types from previous AI research |
|---|---|---|
| **Algorithmic Bias (Bias in Model & Learning Process)** | T2 Classification algorithms<br>T4 Neural network parameters<br>T14 Predictive modeling<br>T15 Reinforcement learning<br>T17 Unsupervised clustering | Measurement bias, Aggregation bias, Evaluation bias |
| **Application-Specific Bias** | T7 Cybersecurity applications<br>T9 Medical applications<br>T10 Information retrieval<br>T13 Image segmentation<br>T16 NLP<br>T18 Computer vision | Deployment bias, Evaluation bias, Aggregation bias |
| **Data Bias (Issues in Dataset & Representation)** | T1 Domain adaptation<br>T6 Bias shift<br>T8 Systematic preference<br>T11 Distribution of training data | Representation bias, Measurement bias, Historical bias |
| **Social and Ethical Bias** | T3 User behavior analytics<br>T5 AI bias, fairness and discrimination<br>T12 Ethical issues | Historical bias, Representation bias, Deployment bias |

**Table 5**
Five topics constituting the algorithmic bias theme.

| ID | Topic | Research Area: | 'Bias' understood as: |
|---|---|---|---|
| T2 | Classification Algorithms | Developing an AI learning method that categorizes unseen data based on seen data. | An inherent element of unbalanced datasets that needs to be accounted for when training classification algorithms. |
| T4 | Neural Network Parameters | Optimizing systems that use Neural Network Learning by manipulating parameters. | A parameter that can be adjusted to improve the performance of Neural Network models. |
| T14 | Predictive Modeling | Optimizing models that are developed to make predictions. | The tendency of predictive models to make systematic errors when calculating, resulting in discrepancy between predictions and actual values. |
| T15 | Reinforcement Learning | Developing models that use reward-based learning. | An element of Reinforcement Learning that is present in the data being used to train models and the nature of how they are trained. |
| T17 | Unsupervised Clustering | Developing systems that automatically detect trends and groupings in data. | An element of clustering that can be, among other things, either an unwanted byproduct or a beneficial tool to enhance performance. |

### 3.2.2. Topic 4: neural network parameters

Research under this topic primarily focuses on developing algorithms for training classical artificial neural networks (ANNs), with specific emphasis on optimizing weight and bias parameters between neuron pairs in multilayer perceptrons. Bias is understood as a parameter of a model that can be adjusted to improve performance. Typically, weights and biases are randomly selected during the optimization process but can lead to improved prediction accuracy on minority classes (Wu et al., 2017). Researchers have proposed advanced methods for finding optimal weights and biases, minimizing a loss function such as a genetic algorithm that incorporates bias directly (Erzurum Cicek & Kamisli Ozturk, 2021) or a combination of metaheuristic algorithms (Ncir et al., 2022).

Most research related to ANN is highly technical and focused on the details of algorithm development. In this context, bias is not necessarily viewed as a negative attribute to be removed, but rather as an essential parameter that can be fine-tuned to improve the performance of the model. An emphasis on parameter optimization strategies reflects a growing recognition of the benefits of introducing bias as a tunable statistical parameter rather than an undesirable issue.

### 3.2.3. Topic 14: predictive modeling

The topic focuses on the role of bias in predictive models and defines bias as the tendency of the model to make systematic errors when predicting outcomes. This notion of bias can be measured based on the difference between the predicted values and the actual outcomes and stems from "erroneous data as inputs in the prediction process [that] may produce inaccurate predictions" (Islam & Hassanzadeh Amin, 2020, p. 1). In other words, bias here represents the degree to which the model predictions diverge from the actual values in real-world applications. The literature centers on AI predictor systems, but the focus on how bias relates to them varies. Some are concerned with making them more interpretable and ensuring that users can identify bias (ElShawi et al., 2021) while others are more technical and are focused on dealing with systems that are better trained to deal with high levels of bias in data (Adnan, 2022; Islam & Hassanzadeh Amin, 2020). The significance of bias in predictive modeling systems is evident, although the viewpoint on how it integrates may vary depending on the developmental phase.

This heterogeneity of arguments under this topic may highlight a fundamental tension between accuracy and reliability in predictive modeling. As AI systems become more sophisticated, the challenge of bias mitigation is evolving from a purely technical problem to a multifaceted issue that intersects with interpretability and practical application. This shift suggests a growing recognition that simply improving model performance is insufficient; there is an emerging need for AI systems that can not only predict accurately but also explain their reasoning and adapt to biased or imperfect data in real-world scenarios.

### 3.2.4. Topic 15: reinforcement learning

The articles in this topic are focused on how bias may manifest itself in reinforcement learning (RL). This topic highlights the significance of bias as a consideration in the development of RL agents, and how researchers could recognize and address these various forms of bias to develop more effective RL algorithms. Bias can take on many forms in this context, including the inductive bias that is inherent to RL models (Chenu et al., 2022) and the tendency for RL models to favor action over inaction (Chen et al., 2020). Additionally, AI researchers discuss the importance of encoding and balancing various learning biases, such as best-response, cautious, and optimistic learning biases, into systems to make RL agents learn more effectively and make better compromises (Crandall & Goodrich, 2011). Bias in this topic is often formulated as model bias, where the data used for training is less robust than actual data, which can be costly to gather and evaluate (Li et al., 2019).

The various forms of bias identified by researchers all contribute to potential inaccuracies in RL models. These biases can lead to suboptimal decision-making, limited generalizability, and a disconnect between the RL model's performance in training versus real-world scenarios.

Researchers highlight the importance of addressing underlying biases in RL systems before they are implemented, demonstrating an understanding of a sociotechnical approach to bias.

#### 3.2.5. Topic 17: unsupervised clustering

This topic investigates how bias arises within unsupervised clustering algorithms, which group data points based on similarity measures to reveal underlying patterns in unlabeled datasets. Unlike supervised models, clustering techniques are designed to assist in exploratory data analysis. In this context, bias can emerge either as an unintended byproduct of the algorithmic process or as a deliberately introduced mechanism to enhance clustering performance. Bias is typically conceptualized in two main ways. First, it can be a harmful artifact, introduced when algorithms misinterpret proximity or similarity in the data, for example, when they fail to distinguish subtle variations between overlapping clusters, which leads to inaccurate representations. Second, bias can be strategically embedded, especially when aligned with domain-specific heuristics, to guide clustering outcomes more effectively.

A typical paper is Gharghabi et al. (2018), who critique standard distance measures, such as Euclidean distance, for their susceptibility to structural biases. These measures often assume consistent data scale and distribution, which does not hold in more complex or noisy time series data. To address this, the authors propose a novel distance metric that better captures local similarity and is more robust to data variation. Researchers in this area have questioned the neutrality of clustering decisions when common heuristics embed implicit assumptions about data structure. In response, newer methods allow for the incorporation of prior knowledge or bias-informed parameters to steer the clustering process toward more meaningful groupings, especially in domain-sensitive contexts such as bioinformatics or social network analysis.

Thus, research in this topic treats clustering bias mainly as a methodological issue. While it may arise unintentionally through flawed distance metrics or simplistic heuristics, it can also be intentionally leveraged to refine clustering performance. Researchers increasingly recognize that improving unsupervised models requires not only better algorithms but also a critical understanding of the assumptions and biases those algorithms encode.

### 3.3. Application-specific bias: topics 7, 9, 10, 13, 16, 18

The application-specific bias theme consists of six topics (see Table 6). Like the topics grouped under the theme of algorithmic bias, the topics featured here mostly share a technical understanding of bias in different application areas such as cybersecurity, medical applications and information retrieval.

**Table 6**
Six topics constituting the application-specific bias theme.

| ID | Topic | Research Area: | 'Bias' understood as: |
|---|---|---|---|
| T7 | Cybersecurity Applications | Optimizing AI Systems that detect and prevent malicious activity in computer networks. | Either as a parameter that can be optimized or as an unwanted element in data that can affect the accuracy of attack detection. |
| T9 | AI Medical Applications | Developing and implementing AI systems in medical settings. | Elements within an AI model that can adversely affect the outcome, potentially hindering its implementation in a medical setting. |
| T10 | Information Retrieval | Developing AI models that are used to retrieve information. | Either as a parameter that can be optimized to improve performance or as an unwanted characteristic of the data used to train models. |
| T13 | Image Segmentation | Correcting bias field" and reducing noise in image segmentation. | Distortion that corrupts MRI images. |
| T16 | NLP | Developing models that utilize natural language processing to make predictions and emulate human behavior. | An element in the NLP model development process that arises from the nature of the training data. |
| T18 | Computer Vision | Developing models that identify objects in images and videos. | A feature of computer vision models that is integrated to more accurately model human behavior when recognizing elements in an image. |

#### 3.3.1. Topic 7: cybersecurity applications

This topic addresses the use of AI, particularly ML and neural networks, for detecting malicious activity in computer networks. The research is situated at the intersection of network security and algorithm optimization, with a strong focus on improving the accuracy and performance of intrusion detection systems. The typical methodological approach involves the development and tuning of AI-based classifiers that can distinguish between normal and abnormal network traffic.

In this topic, bias is conceptualized in two main ways: (1) as an unwanted distortion in data that can impair the accuracy of attack detection (e.g., class imbalance in training datasets), and (2) as a tunable parameter within AI models that can be strategically optimized to enhance detection accuracy. This dual understanding reflects a domain-specific and performance-oriented conception of bias.

Moghanian et al. (2020) exemplify the technical focus of this topic. Their paper introduces an intrusion detection system (GOAMLP) that integrates a multilayer perceptron neural network with the Grasshopper Optimization Algorithm (GOA). In this method, bias parameters (along with weights) are not seen as problematic artifacts but as levers for improving model performance. GOA is used to iteratively optimize these parameters, minimizing classification error and improving detection accuracy. The study relies on benchmark datasets like KDD and UNSW, demonstrating performance gains over other methods, such as XGBoost and other metaheuristic algorithms.

Other studies within this topic define bias more traditionally, for example, as a result of imbalanced datasets where minority attack classes are overlooked. Papamartzivanos et al. (2018) propose heuristic methods to compensate for this skew. Limthong et al. (2015) discuss temporal biases in traffic (e.g., favoring new over old traffic) and how weighting strategies can improve detection, while Gupta and Alam (2023) use optimization to refine bias parameters in forgery detection tasks. These examples show how bias is not universally conceptualized as a defect to be eliminated, but also as a design feature to be tuned. This reflects a utilitarian ethos in AI security research: if bias tuning can improve detection rates, it is embraced.

#### 3.3.2. Topic 9: AI medical applications

The research presented in this topic focuses on evaluating the viability of AI algorithms before they are implemented in a medical setting and specifically investigates these systems for risk of bias. Bias here typically refers to "factors that can systematically affect the observations and conclusions of the study and cause them to be different from the truth" (Higgins et al., 2011). Bias in medical applications of AI is a pressing issue, and concerns about the development, validation, and reporting of AI algorithms have emerged, including whether researchers are adequately auditing AI systems that assist in diagnostic medicine and patient management (Suri et al., 2022). Bias is seen as a negative aspect of an AI system and needs to be mitigated or removed before the system is put into practice in real-world medical applications. However, researchers emphasize that the majority of AI models have yet to be implemented in actual clinical settings. They argue for the necessity of improving methodologies, reporting, and validation regarding the bias problem before these models can be effectively deployed on a large scale

(Silva et al., 2020).

In research related to healthcare, bias is typically framed as a systemic deviation that can compromise the validity and reliability of AI systems, potentially leading to discrepancies between algorithmic outputs and clinical goals. Researchers in this domain approach bias with heightened vigilance, emphasizing the need for comprehensive auditing and validation protocols before deploying AI systems in diagnostic medicine and patient management. This stance reflects an awareness of the high stakes involved in medical applications where algorithmic bias could directly impact patient outcomes.

#### *3.3.3. Topic 10: information retrieval*

This topic focuses on the role of bias in the design and evaluation of information retrieval (IR) systems, particularly in tasks such as document ranking, query expansion, and user behavior modeling. The IR research here is driven by efforts to enhance retrieval effectiveness, often through statistical tuning or algorithmic refinement. However, it also reveals competing notions of bias. Some studies treat bias as a performance-enhancing variable, while others approach it as a source of distortion stemming from noisy or inconsistent user interaction data.

Bias in this topic is conceptualized in two ways. First, it is seen as an intentional and adjustable component of system design, for example, in result diversification, query expansion, or term weighting, where elements of bias are introduced to improve retrieval outcomes. Second, it is treated as an unwanted artifact of data, such as click behavior or user feedback signals, which can lead to inaccurate relevance estimation and impaired system evaluation. This duality characterizes the pragmatic approach to bias taken within IR research.

An example is provided by Lam-Adesina and Jones (2001), who investigate the use of document summaries for term selection in pseudo-relevance feedback (PRF). They argue that standard PRF approaches may introduce irrelevant expansion terms due to noise in full-text documents and propose summary-based methods to reduce such distortion. Their experiments show that biasing summaries toward the query significantly improves term selection and retrieval effectiveness. Here, bias is actively engineered into the system. It is treated not as a flaw, but as a targeted design feature to guide query refinement.

At the same time, IR researchers remain attentive to user behavior data biases, especially in click-through logs and query reformulations. Studies highlight that click-based relevance signals can vary by user, topic, or interface position, leading to challenges in evaluating system performance objectively (Scholer et al., 2008). Such work frames bias as a contaminant that must be carefully managed or mitigated.

Thus, the topic illustrates a tension between operational and epistemic roles of bias. On one hand, IR systems benefit from the strategic use of bias to improve output quality. On the other hand, they are vulnerable to bias in training and evaluation data, particularly from user interactions that introduce noise and inconsistency.

#### *3.3.4. Topic 13: image segmentation*

The research presented in this topic addresses a specific type of bias that affects the field of magnetic resonance imaging (MRI). This bias is known as the "bias field," which is a low-frequency and very smooth signal that corrupts MRI images, particularly in images produced by older MRI machines (Juntu et al., 2005). The proposed solutions are highly technical and involve accounting for spatial information during different processes, such as fuzzy clustering (Ji et al., 2014), using gradient minimization to enforce a piecewise constant solution (Duan et al., 2014), and image segmentation (Li et al., 2011). The bias addressed in MRI training data models is not linked to social contexts or user-related factors, yet it still persists as an underlying issue in the training data that could impact the AI model training. The reliance on old images with high levels of inhomogeneity necessitates the use of statistical methods to tackle the bias.

Researchers are focused on a domain-specific type of bias, which is not socially contextual but is nevertheless seen as an element of AI systems that needs to be systematically addressed. They are focused on developing advanced algorithms to remove the bias field. While this is a technical solution-oriented view of bias, the perception of real-world implications suggests an understanding of how bias permeates through the AI lifecycle and has the potential to negatively impact outcomes.

#### *3.3.5. Topic 16: NLP*

This topic explores how bias emerges in natural language processing (NLP) systems from large, organically sourced text corpora such as novels, online forums, Wikipedia, and social media. NLP systems often mirror human language patterns without accounting for biases embedded in their training data. As a result, bias is understood primarily as a feature of the input data, which reflects existing social inequalities. The research in this area seeks to detect, quantify, and mitigate these biases, often highlighting the harms and risks of deploying NLP models trained on unfiltered text.

In this topic, bias is typically conceptualized as representational bias (i.e., systematic distortions in how people, identities, or social groups are portrayed) and is attributed to skewed or incomplete training data. This includes overrepresentation of dominant perspectives, underrepresentation of marginalized voices, and the reproduction of stereotypes. Because language data is often scraped from the open web or public platforms, these biases are not only common but structurally baked into the data sources.

A representative study is Fast et al. (2016), who analyzed a dataset of 1.8 billion words from an online fiction writing community and applied a stereotype lexicon to quantify gendered linguistic patterns. They found substantial male overrepresentation and clear patterns of gender stereotyping, such as the tendency for male characters to be described in terms of action and danger, while female characters were more often associated with appearance and domesticity. Other studies extend the critique to multilingual corpora. For instance, Chen et al. (2021) examined Wikipedia entries across nine languages and demonstrated variation in both quality and quantity of content. Moreover, Bender et al. (2021) emphasize the risks of relying on "found data" without regard for its provenance, raising questions about whose voices are amplified or excluded in large-scale language models.

Scholars in this topic suggest that mitigating these biases requires both technical interventions (e.g., debiasing algorithms, diverse training sets) and more ethical data sourcing, as well as inclusive representation. They call for more deliberate curation of training data and critical reflection on the social inequality implications of NLP systems.

#### *3.3.6. Topic 18: computer vision*

Computer vision has grappled with the issues associated with "saliency," which refers to "the quality that makes certain regions of an image stand out from the visual field and grab our attention" (Bruno et al., 2020, p. 1). Researchers have proposed methods to introduce various forms of bias to more effectively detect salient aspects of an image or video. For example, DeepVS, as a deep learning approach, uses data gathered from eye-tracking humans to determine how humans have a predisposition towards moving objects. The notion of bias is added to the learning network to better predict saliency (Jiang et al., 2018). Center-bias is another type of bias that is frequently mentioned in this topic. By incorporating the idea that human "gaze fixations are biased toward the center of natural scene stimuli" (Tseng et al., 2009, p. 1) into models that combine the bias with pixel-based saliency (Borji & Tanner, 2016), stereoscopic image detectors (Fang et al., 2014), and in the form of a prior (Borji & Tanner, 2016), have all proven to be more effective than traditional, non-biased computer vision systems. Because researchers are trying to recreate human vision in AI systems, the biases that humans possess are being recreated and are seen as a tool to improve model performance.

In the context of computer vision, bias is increasingly viewed as a multifaceted phenomenon with both positive and negative implications, depending on its origin and application. For example, the bias as a parameter in computer vision systems can be introduced and used to enhance performance. Bias can be thought of as a bridge between human vision and ML, where biases inherent to human attention and cognition, such as the center-bias and motion bias, are intentionally replicated within models to guide the algorithm in predicting salient features of images and videos. By embedding these cognitive shortcuts or tendencies into AI models, researchers aim to narrow the gap between human and machine perception.

### 3.4. Data bias: topics 1, 6, 8, 11

The data bias theme consists of four topics (see Table 7). In comparison with the topics grouped under the themes of algorithmic bias and application-specific bias, the topics featured here have a more sociotechnical understanding of bias, combining technical and social elements.

#### 3.4.1. Topic 1: domain adaptation

This topic centers on the adaptability of systems across domains. Domain adaptation is key in training AI systems because the performance of models may often deteriorate when applied to different but adjacent domains. The desire to incorporate previously unseen classes into models that span domains has led researchers to adjust models trained on data from one domain to different domains. This topic acknowledges that models tend to be biased between seen and unseen classes and presents methodologies and technical solutions to solve it. Techniques, such as zero-shot learning that aim to "recognize unseen samples that never appear during training" have been explored (Yang et al., 2023, p. 281), with model performance and accuracy being important metrics for determining their viability. Researchers also explore embedding semantic relationships (e.g., Zhao et al., 2022), combining supervised and unsupervised learning (e.g., Balgi & Dukkipati, 2022), and augmenting training samples with confident target unseen samples (e.g., Li et al., 2020) to reduce the bias that is inherent in domain adaptation.

The bias discussed in this topic is rooted in the difference between the source and target domains of the model. Mitigating this sort of bias requires training models that can generalize across domains, for instance, by learning transferable representations. It can also mean striking a balance between overfitting to the source domain and underfitting the target domain. In other words, domain adaptation methods such as transfer learning can improve the performance of AI models, but they require more training to accommodate the diversity of unseen domains (Balgi & Dukkipati, 2022).

**Table 7**
Four topics constituting the data bias theme.

| ID | Topic | Research Area: | 'Bias' understood as: |
|---|---|---|---|
| T1 | Domain Adaptation | Adapting models trained on one domain to perform as effectively on related but different domain data. | An inherent element of model development that occurs when models are utilized on different, but related data domains. |
| T6 | Covariate Shift | Improving the training of Convolutional Neural Networks. | A technical aspect of learning models that occurs during training and can impede both learning speed and accuracy. |
| T8 | Systematic Preference | Optimizing evolutionary and stochastic local search algorithms. | A parameter that can be introduced to better represent the problem the algorithm is trying to solve. |
| T11 | Distribution of Training Data | Enhancing the underlying data that is used to train AI systems. | A broad range of parameters that impact the quality of data used to train models, including skewed labels, improper data distribution, and data noise. |

#### 3.4.2. Topic 6: bias shift

This topic focuses on improving the performance of convolutional neural networks (CNNs) by developing innovative methods to mitigate bias shift. CNNs are popular deep learning algorithms in tasks such as image and facial recognition that are subject to bias shift during training (Sakai, 2020). Bias shift occurs when weights are updated and can impede learning speed and accuracy (Clevert et al., 2016). The technical focus of this topic revolves around developing innovative methods to mitigate bias shift and thereby enhance CNN performance, particularly for tasks such as image and facial recognition. Bias shift occurs during the weight update process in training and can significantly impede both the learning speed and overall accuracy of the model.

This topic alludes to a rather singular and specific view of bias, and the studies are geared towards optimizing models using methods such as modifying activation functions that use rectified linear units (ReLUs) (Zhao et al., 2018). While these techniques offer valuable improvements in CNN efficiency and accuracy, they represent a narrow, technically focused perspective on bias in machine learning systems. As such, when describing bias, researchers in this context often pay less attention to application and domain specificity, or they may not consider the real-world outcomes or implications of the algorithm.

#### 3.4.3. Topic 8: systematic preference

This topic focuses on optimization algorithms, particularly evolutionary and stochastic local search algorithms, where bias is introduced as a form of systematic preference to guide search processes. In this context, bias is not understood as a distortion to be mitigated but rather as a computational strategy that helps algorithms converge more efficiently toward optimal solutions. Researchers frame bias as a parameter that can be designed and tuned to reflect the structure of the underlying problem space, improving performance in solving complex optimization problems.

Fu et al. (2022) exemplify this approach through their work on satisfiability problems. They propose an improved selection probability mechanism that incorporates bias into the random walk strategy used in local search. Rather than relying on a standard, uniform selection process, they implement a biased random walk that preferentially guides the algorithm toward more promising solution directions. Their results show improved solution accuracy and faster convergence, highlighting how bias can serve as an optimization tool.

Bias is thus understood as a form of intentional asymmetry introduced into algorithmic processes. For example, biased mutation operators in genetic algorithms or biased neighborhood selection in local search are used to focus computational resources on more likely successful regions of the search space (Grasas et al., 2017). These methods are typically evaluated in terms of efficiency and success rate rather than ethical implications or application-specific outcomes. Researchers in this area are often not concerned with the final use of the algorithm but rather with the internal logic and performance of the optimization process itself.

#### 3.4.4. Topic 11: distribution of training data

This topic addresses forms of bias that affect the quality of training data used in AI systems, particularly annotated or crowdsourced datasets. The focus is on identifying and correcting issues such as distributional bias, labeling noise, and biased annotations, which can collectively undermine model generalizability and accuracy. Researchers in this area stress the importance of pre-modeling data quality and propose technical solutions to improve training data before algorithmic development begins.

Bias in this topic is broadly conceptualized as any structural flaw in the training data that can distort downstream model behavior, including skewed class distributions, subjective labeling tendencies, and annotation inconsistencies. Such biases are often introduced during the data collection and labeling phases and can propagate through the entire AI pipeline if left unaddressed.

Wu et al. (2022) is a typical study. They develop a method called Label Aggregation with Clustering for Biased Labeling (LACBL), which targets biased crowdsourced labeling, where annotators disproportionately favor one class over another. LACBL first applies clustering techniques to detect latent label distribution imbalances, estimates the degree of labeling bias by comparing inferred and observed label distributions, and adjusts the label set accordingly. Across several real-world datasets, LACBL shows that detecting and correcting for bias prior to aggregation can significantly improve label quality and downstream model accuracy. Other studies offer additional techniques to mitigate data distribution issues. For instance, Hu et al. (2016) propose a Bayesian approach to integrate the outputs of individual labelers, treating each as a separate classifier to address inconsistencies. Wang et al. (2017) introduce a method for measuring data representativeness and discriminativeness using semi-supervised active learning, improving the selection of training examples.

The literature within this topic frames bias as a foundational issue in ML, which must be addressed at the data level, rather than deferred to model interventions. This proactive stance highlights an evolving recognition that data quality is not just a technical concern but a prerequisite for building trustworthy AI systems.

### 3.5. Social and ethical bias: topics 3, 5, 12

The social and ethical bias theme consists of three topics (see Table 8). Like in the previous theme of data bias, the understandings of bias here tend to be sociotechnical. Overall, this is the theme most concerned with the social implications of bias in real-life settings, especially the notion of discrimination.

#### 3.5.1. Topic 3: user behavior analytics

This topic focuses on introducing technical methods to deal with biases that arise from user behavior data in recommendation systems. Researchers within this topic are developing innovative strategies to enhance the technical accuracy of these models. Bias can permeate user behavior analytics systems in the form of popularity bias or review-rating mismatch (Aralikatte et al., 2018). It can be understood here based on problems, such as the prevalence of nonconsecutive, sampled user data being used to predict mobility patterns and behavior in location-based social networks (LBSNs) (Jin et al., 2016). This topic specifically addresses the inconsistency between user rating scores and the sentiment of their reviews on platforms like Amazon (Aralikatte et al., 2018; Xu et al., 2019) and the vulnerability of collaborative recommender systems to malicious users who inject fake data, potentially biasing the output (Zhang et al., 2009).

Biases can result in mispredictions of user behavior and preferences; researchers in this topic are concerned with the implications of depending solely on this data. These biases often arise from the feedback loops of user behaviors (Xu et al., 2019). As noted, for example, popularity bias in recommendation algorithms can create a self-reinforcing cycle where popular items become even more visible, potentially stifling diversity and innovation. As another example, the mismatch between review content and numerical ratings or the use of incomplete or non-representative user data (e.g., sampled location data in social networks) presents another challenge by misrepresenting the user sentiment and, therefore, generating skewed recommendations.

**Table 8**
Three topics constituting the social and ethical bias theme.

| ID | Topic | Research Area: | 'Bias' understood as: |
|---|---|---|---|
| T3 | User Behavior Analytics | Leveraging user behavior data, such as online reviews or location data, to train prediction systems. | An unwanted consequence of using behavior data to train prediction models. |
| T5 | AI Bias, Fairness & Discrimination | Mitigating discrimination and raising fairness in training data and algorithmic decision-making. | Discrimination based on sensitive attributes in the data used for algorithm creation and implementation. |
| T12 | AI Ethical Issues | Addressing ethical and social consequences of bias in AI (e.g., diversity, inclusion, and accessibility). | Consequences of AI systems that typically emerge during the implementation of algorithms in society. |

#### 3.5.2. Topic 5: AI bias, fairness, and discrimination

This topic centers on the presence of algorithmic bias in AI models and the strategies that can be used to alleviate it. Bias here can be understood as "discrimination due to sensitive attributes such as gender, race, or ethnicity" (Ouadrhiri & Abdelhadi, 2021, p. 1). To address this issue, researchers have proposed various technical approaches with the goal of maintaining performance while achieving fairness (Hu & Rangwala, 2020). Introducing a randomized response mechanism during pre-processing the dataset to mitigate the inequity (Ouadrhiri & Abdelhadi, 2021) or constructing a novel face image dataset that is balanced on race (Kärkkäinen & Joo, 2021) are examples of solutions that deal directly with datasets and are designed to account for bias when developing models. Model-agnostic solutions, such as OmniFair (Zhang et al., 2021), allow practitioners to add user-specified fairness constraints to the optimization process of a model. The importance of mitigating bias is emphasized in this topic because of a shared belief that AI models are increasingly used in high-stakes applications.

This topic addresses not only the manifestation of bias in ML models as discrimination based on sensitive attributes such as gender, race, or ethnicity but also develops technical solutions to balance fairness and performance metrics. Research under this topic has explored a range of approaches, from data preprocessing methods to model-agnostic optimization techniques. The technical challenges of addressing algorithmic bias highlight the ongoing need for research in areas like fairness-aware ML, causal inference for bias detection, and adaptive learning algorithms that can adjust dynamically to changing fairness standards.

#### 3.5.3. Topic 12: AI ethical issues

This topic outlines the ethical implications and social consequences of using AI systems in practice. Bias, in this context, is seen as an "inherent" (Avellan et al., 2020), social issue that can impact different groups, such as youth (Forsyth et al., 2021), marginalized groups (Tachtler et al., 2021), and those involved in the criminal justice system (Malek, 2022). Articles in this topic explain the negative consequences of using AI in specific contexts and highlight the need to oversee the development, deployment, upkeep, and decommissioning of AI systems (Solomonides et al., 2022) as well as the need to develop strategies for effective AI audit (Eitel-Porter, 2021). Furthermore, they make a case for increased transparency in AI systems to mitigate issues such as biased outcomes (Maclure, 2021). Focusing on the real-world impacts of AI systems, the research here not only amplifies the concern that failing to detect and address bias in these systems could result in harm to various communities but also devises solutions to enhance the safety and explainability of AI. However, it is unclear how widespread this understanding of bias is, how it affects the technical development process, and whether the AI research community views this understanding as ground truth.

The research under this topic frames bias not merely as a technical issue to be solved but as a complex social problem that disproportionately affects marginalized groups. Researchers emphasize a holistic approach to addressing AI bias, advocating for comprehensive oversight throughout the entire AI lifecycle and highlighting the critical importance of transparency and effective auditing mechanisms. This perspective represents a significant shift from purely technical considerations to a broader understanding of AI's societal impacts, where bias is viewed through the lens of potential harm to the end-users or even non-users in the real world.

## 4. Discussion

### 4.1. Summary and theoretical implications

While bias has received ample attention in AI research and public discourse recently (Kordzadeh & Ghasemaghaei, 2022; Mehrabi et al., 2022), our comprehensive CLR provides a novel and unique contribution. It goes beyond existing attempts to synthesize the literature by covering a much larger and broader body of texts (6520 entries), showing how the AI research community acts as an important ethical agent (Griffin et al., 2023). Our analysis revealed that this attention to bias follows different conceptual and methodological commitments rather than a unified understanding. Using topic modeling, we identified 18 topics that span various AI research areas and issues but show certain internal consistency. These 18 topics can be grouped into four broader themes, which partly align but also go beyond existing typologies of bias (Mehrabi et al., 2022; Schwartz et al., 2022; Suresh & Guttag): algorithmic bias, data bias, application-specific bias, and social and ethical bias. The four themes can be subsumed under two cross-cutting conceptions: technical and sociotechnical. Algorithmic and application-specific biases primarily follow the technical conception, while data bias, along with social and ethical biases, aligns more with the sociotechnical side. However, this mapping is not rigid, as many topics blend both conceptions (e.g., some research in the NLP topic uses a sociotechnical conception), revealing definitional tension rather than neat separation. The following discussion of these two main conceptions is therefore a relatively abstract synthesis of the findings.

On the one hand, the *technical* conception of bias characterizes bias in AI systems as a persistent divergence from accuracy, evident in the outputs of an algorithm. This technical conception focuses on operationalizing bias as a statistical problem or the computational performance of the AI model that can be captured through clear and often specific variables. Accordingly, this type of bias is sometimes called statistical bias (Barocas & Hardt, 2023). Specifically, this line of research describes and proposes mathematical and statistical techniques to mitigate biases in the preprocessing, in-processing, and post-processing phases of data pipelines for AI models (Ferrara, 2023). Moreover, it includes approaches that fine-tune models to minimize both statistical and computational aspects of bias. Consequently, the primary goal is to enhance the performance of AI systems as technological systems.

On the other hand, the *sociotechnical* notion of bias is about the social consequences of algorithm design, for example, systematic and unfair outcomes of the systems. It captures the interplay of technical and social aspects (Sarker et al., 2019). From this perspective, bias can be defined as a deviation from equality or fairness of outcomes relative to different social groups. The concept of bias is understood as a social and contextual phenomenon, relating to social and ethical issues of discrimination, unfairness, and social injustice (Fosch-Villaronga et al., 2021; Williams et al., 2018). Algorithms do not perform in a vacuum but are integrated into biased institutions and social contexts shaped by existing power relations and structural inequalities (Asatiani et al., 2021; Lutz, 2024; Sawyer & Jarrahi, 2014). They can exacerbate or introduce bias by prioritizing aspects of human behavior that are easily measurable over those that are difficult or impossible to quantify (Ntoutsi et al., 2020; Soprano et al., 2024). As such, while technical and statistical approaches can help mitigate bias in the data and model, they are inadequate in capturing and addressing the social contexts of bias. This second conception of bias is sociotechnical since bias is understood as arising from the interplay between (a) the social context of data, the development and use of algorithmic systems, as well as (b) technological characteristics of these systems (e.g., inherent opacity of deep learning systems, Asatiani et al., 2021). In this broader, contextual definition of bias, social dynamics are considered significant in shaping the development of biased AI systems and the integration of biased AI inferences in organizational processes (Asatiani et al., 2021; Tachtler et al., 2021).

Across the 18 topics, four themes, and two conceptions, our analysis reveals a differentiated research landscape, where multiple meanings and treatments of bias coexist without an overarching framework and understanding dominating. While the diversity of perspectives reflects the complexity of bias as a phenomenon and the growth and diversification of the AI research community, it also makes it difficult for AI researchers to engage in cross-cutting conversations or develop coherent responses to bias in AI. Much of the technical literature treats bias as an intrinsic property of algorithms or data, failing to recognize how bias manifests differently across application domains, social settings, and deployment environments.

Our review also uncovers several problematic assumptions regarding the nature and role of bias. A persistent assumption is that unbiased or neutral AI systems are achievable (Dwork et al., 2012), overlooking the fundamentally value-laden nature of technology design and deployment (Cave, 2020). Friedman and Nissenbaum (1996) argued in their foundational research that all computer systems embody particular values and perspectives. The pursuit of neutrality can be misleading; the more productive question is not how to eliminate bias entirely but how to ensure that AI systems reflect and promote values such as fairness, equity, and social good (Soprano et al., 2024).

Another common assumption is that algorithmic bias can be addressed in isolation from broader social inequalities pervasive in organizations or society at large. This is evident in work that focuses narrowly on bias mitigation techniques without attending to the social, institutional, or historical conditions in which bias arises or manifests (e.g., Bolukbasi et al., 2016). Many technical approaches treat bias as a discrete, internal feature of AI systems rather than a reflection of entrenched social inequalities that shape data, design processes, and institutional adoption. This narrow framing leads to surface-level fixes that fail to address some root causes of algorithmic bias, discrimination, and harm. As Tachtler et al. (2021) and Griffin et al. (2023) note, bias mitigation efforts must be complemented by a broader engagement with the social and ethical responsibilities of AI developers and the systems they create.

### 4.2. Research implications

Based on our findings, AI researchers should clarify their approach towards bias and its role, origins and limitations within their work. This discussion can be framed along several dimensions, including 1) the nuances in their understanding of bias (e.g., seeing bias as a challenge or an opportunity; as error or discrimination – or both), 2) the perceived origin of the bias, 3) its link to the data that drives the AI model, 4) how it manifests (e.g., as a statistical error versus discrimination against certain groups), 5) its visibility and detectability (e.g., whether the bias is easily observable in the system's outputs or is subtler and harder to quantify), 6) its social and ethical implications, 7) its ties to human behavior and context, 8) the stakeholders it may affect, and 9) strategies for remediation.

The four themes provide complementary entry points for future research. Studies on *algorithmic bias* should expand beyond fairness metrics to explore real-world impacts, trade-offs, and unintended consequences of model tuning. Research into *application-specific bias* would benefit from domain-aware evaluation frameworks and situated testing practices. Work on *data bias* could prioritize participatory dataset design, bias documentation standards, and tools for tracing bias through the data pipeline (Jarrahi et al., 2023b). Finally, *social and ethical bias* calls for deeper integration of legal, political, and community perspectives, especially in high-stakes contexts.

A sociotechnical understanding can help develop an interoperable language that reframes bias in the context of a wider AI interdisciplinary discourse (Bechky, 2003). AI researchers need not abandon their distinct interpretations of bias, but following Sarker et al. (2019), who advocate for the sociotechnical perspective as an axis of cohesion for the information systems discipline, a process of exchange can bridge gaps between technical, ethical, and humanistic viewpoints, ultimately fostering more accountable and transparent AI implementation (Buhmann & Fieseler, 2023).

Mobilizing visions such as human-centered AI or interdisciplinary publication platforms can serve as coordination opportunities for AI researchers across disciplines (e.g., Shneiderman, 2022). Additionally, future work should examine how bias is operationalized in practice, for example, how medical or cybersecurity research treats bias as both a liability and a resource. These tensions reveal opportunities for clearer guidelines on when and how bias can be harnessed to improve model performance without compromising fairness.

Moreover, some topics within the medical, cybersecurity, and computer vision domains point to a growing tension between bias as a tool and a challenge. The AI research community should explore how operationalizing bias parameters (in models or images) can be transformed into an asset for enhancing AI performance while ensuring that transparency and fairness standards are upheld. Lastly, the continued evolution of AI systems means that bias must be considered and addressed throughout the whole model lifecycle. This calls for comprehensive auditing frameworks and real-time bias detection mechanisms that are adaptable to shifting values, societal expectations, and stakeholder needs (Sandvig et al., 2014).

### 4.3. Managerial implications

This research provides insights into the development, implementation, and oversight of AI systems in organizations, five of which are discussed in the following. First, AI practitioners and managers should foster *cross-functional* dialogue to align interpretations of bias for both ethical and business reasons**.** Organizations stand to benefit from proactively engaging with the concept of bias beyond regulatory compliance or ethical responsibilities alone. Clear discussions and systematic structures for understanding the complexity of AI bias and addressing it can enhance organizational effectiveness by improving decision-making quality, increasing trust and acceptance of AI-empowered decisions among stakeholders, and reducing the risk of costly reputational damage. Systematic organizational efforts to address biases in AI processes can improve the robustness, reliability, and generalizability of AI outcomes. As AI systems take on more important roles in decision-making, contextualizing bias within specific organizational frameworks helps align them more closely with strategic objectives, enhance user experience, and deliver outcomes that are more effective across diverse user groups. As such, organizational proactive engagement with bias is an increasingly strategic investment that contributes positively to organizational agility, innovation capacity, and overall operational excellence.

Realizing this vision requires alignment and shared understanding among multiple organizational stakeholders. Our findings, however, make it clear that bias in AI is currently understood differently across technical, data, domain-specific, and integration teams. Managers should actively create organizational structures, such as cross-functional working groups, that bring together AI developers, data scientists, ethicists, legal advisors, and operational staff to share their varied understandings of bias and collaboratively develop solutions to remedy bias in AI as a multidimensional issue (Buhmann & Fieseler, 2023).

Second, those involved in AI development and deployment should *integrate both technical and contextual considerations into AI governance*. Our analysis demonstrates that technical solutions to bias (e.g., improved data sampling, model adjustments, or algorithmic fairness metrics) are helpful but often insufficient. Organizations developing or deploying AI must ensure that technical teams are also attuned to how bias manifests differently in various real-world application contexts. Therefore, AI governance frameworks should include context-sensitive risk assessments that account for model performance metrics and contextual and social impacts (Felzmann et al., 2020).

Third, bias should be treated as a *dynamic issue requiring ongoing monitoring and organizational learning*. As both a technical and contextual issue, bias does not end at model deployment. Managers should invest in mechanisms for post-deployment monitoring, feedback loops, and periodic bias audits, particularly for AI systems integrated into high-stake decision-making. These mechanisms should involve technical error detection, feedback from multiple stakeholders, and assessment impacts that reflect how bias evolves over time and downstream in dynamic organizational and social contexts (Raji et al., 2020).

Fourth, AI practitioners should *recognize and confront the value-laden nature of AI systems*. As noted earlier, a problematic assumption has to do with the belief that bias can be fully removed to achieve “neutral” AI systems. In reality, all AI systems inexorably embed certain value systems and priorities (Friedman & Nissenbaum, 1996). Thus, the practical conversation should shift from bias elimination to value alignment, asking: whose values are represented and reinforced in the AI system? Another important step is engaging various stakeholders early in design and decision-making processes. In this context, participatory practices that include impacted communities, such as frontline users or public interest advocates in design and governance discussions, can help bring hidden assumptions to the fore and promote more equitable AI adoption outcomes.

Finally, there should be increased *investment in capacity building and reflexive training*. Given the varied and often implicit assumptions about bias held by different teams and professional backgrounds, organizations should invest in continuous learning opportunities on technical bias mitigation techniques, ethical reasoning, sociotechnical systems thinking, and inclusive design (Madaio et al., 2020). These training programs should support employees in thinking beyond static compliance checklists and toward deeper, reflexive engagement with the implications of AI systems in their respective work contexts.

### 4.4. Limitations and future research

This research has limitations that point to future research opportunities. One area of improvement is the reliance on the title, keywords, and abstract for data. While those elements, especially the abstract, offer a concise summary of research, examining the full text could provide more comprehensive insights. Comparing the information captured in abstracts versus full-text analysis could enhance methods of mining literature reviews for information on bias. Additionally, this study employed LDA, a standard method in literature analysis. Future research could explore alternative topic modeling approaches, which may reveal more complex relationships between topics. While our analysis comprehensively covered a broad range of computer science outlets to capture thematic diversity and methodological rigor, we acknowledge that it did not include certain venues such as PubMed or medical AI conferences like MICCAI. This exclusion may limit the extent to which our findings reflect the nuances of AI bias in clinical and healthcare contexts. However, as noted, by integrating discipline-agnostic databases such as WOS, we still covered a substantial body of literature. Nevertheless, future research should extend the scope by incorporating these databases and outlets to provide a more holistic view of AI bias across diverse application domains. Moreover, our search was restricted to English-language publications, which may have excluded relevant research in other languages. Future research should extend the scope by incorporating diverse databases, outlets, and linguistic contexts and by integrating complementary bibliometric methods such as citation network analysis (Waltman & van Eck, 2012) to provide a more holistic view of AI bias across domains. Finally, a potential issue of our approach - and of topic modeling more generally - is that the topics do not center fully on the concept of interest, namely bias, but reflect other logics in the corpus. We tried to address this by strongly leveraging human oversight, especially in the topic analysis (see sub-section 2.3 Topic

Analysis), and made sure that our 18 topics, four broader themes and two cross-cutting conceptions reflect distinct understandings of and approaches to AI bias. Nevertheless, future research is encouraged to conduct a thorough and comprehensive review of publications that focus exclusively on bias in AI research. Given the breadth and interdisciplinarity of the topic, a scoping review approach would be a meaningful method to do so (Levac et al., 2010).

## 5. Conclusion

How the research community, as a key stakeholder, defines, makes sense of, and addresses AI bias has implications for managing such bias. The issue of bias has been studied for many years, and there has been a growing interest and attention paid to the problem (Ali et al., 2024). However, there remains a lack of consensus on what bias actually means. Given the growing importance of biased decision-making by AI systems, it is important to engage more explicitly with the concept across various sub-communities. The challenges presented by bias require a multidisciplinary perspective and ongoing dialogue between these sub-communities, as well as with other disciplines that can help contextualize the issue. While the problem of bias can be understood and mitigated through technical, computational, and statistical mechanisms, we argue that the concept cannot be reduced to a mere statistical or technical definition. Instead, it encompasses complex, dynamic, social, and often domain-specific issues that cannot be fully addressed solely by these technical operationalizations.

## CRediT authorship contribution statement

**Mohammad Hossein Jarrahi:** Writing – review & editing, Writing – original draft, Visualization, Validation, Software, Resources, Methodology, Investigation, Formal analysis, Data curation, Conceptualization. **Amir Karami:** Writing – review & editing, Writing – original draft, Visualization, Validation, Software, Resources, Methodology, Investigation, Formal analysis, Data curation, Conceptualization. **Patrick Conway:** Writing – review & editing, Writing – original draft, Validation, Investigation, Formal analysis. **Ali Memariani:** Writing – review & editing, Writing – original draft, Validation, Investigation, Formal analysis, Conceptualization. **Christoph Lutz:** Writing – review & editing, Writing – original draft, Validation, Investigation.

## Funding

The last and corresponding author (Lutz) was partly financed by the Research Council of Norway within the research project 299178 *"Algorithmic Accountability: Designing Governance for Responsible Digital Transformations" while working on this article.*

## Declaration of competing interest

The authors declare that they have no known competing financial interests or personal relationships that could have appeared to influence the work reported in this paper.

## Appendix A. Supplementary data

Supplementary data to this article can be found online at https://doi.org/10.1016/j.techsoc.2025.103127.

## Appendix I. Top-5 most frequent conferences and journals per topic.

| Topic | Top-5 Most Frequent | |
|---|---|---|
| | Conferences | Journals |
| **Domain Adaptation** | ● IEEE/CVF Conference on Computer Vision and Pattern Recognition (CVPR)<br>●Lecture Notes in Computer Science, Computer Vision (ECCV)<br>●IEEE/CVF International Conference on Computer Vision (ICCV)<br>●Advances in Neural Information Processing Systems<br>●ACM International Conference on Multimedia | ●IEEE Access<br>●Neurocomputing<br>●Pattern Recognition<br>●IEEE Transactions on Image Processing<br>●IEEE Transactions on Pattern Analysis and Machine Intelligence |
| **Classification Algorithms** | ●International Joint Conference on Neural Networks (IJCNN)<br>●IEEE/CVF Conference on Computer Vision and Pattern Recognition (CVPR)<br>●Lecture Notes in Computer Science, Knowledge Discovery in Databases (PKDD)<br>●IEEE International Conference on Tools with Artificial Intelligence (ICTAI)<br>●IEEE International Conference on Big Data (Big Data) | ●IEEE Access<br>●Expert Systems with Applications<br>●Neurocomputing<br>●Pattern Recognition<br>●Knowledge-Based Systems |
| **User Behavior Analytics** | ●ACM Conference on Recommender Systems<br>●ACM International Conference on Information & Knowledge Management<br>●ACM Web Conference<br>●ACM International Conference on Web Search and Data Mining<br>●International ACM SIGIR Conference on Research and Development in Information Retrieval | ●Expert Systems with Applications<br>●IEEE Access<br>●Applied Intelligence<br>●Data Mining and Knowledge Discovery<br>●Decision Support Systems |
| **Neural Network Parameters** | ●Chinese Control Conference (CCC)<br>●European Conference on Antennas and Propagation (EuCAP)<br>●IEEE Nuclear Science Symposium Conference Record<br>●Lecture Notes in Computer Science, Advances in Natural Computation<br>●International Conference on Frontier of Computer Science and Technology | ●IEEE Access<br>●Neural Computing and Applications<br>●Applied Soft Computing<br>●Engineering with Computers<br>●International Journal of RF and Microwave Computer-Aided Engineering |
| **AI Bias, Fairness & Discrimination** | ●AAAI/ACM Conference on AI, Ethics, and Society<br>●ACM Conference on Fairness, Accountability, and Transparency<br>●International ACM SIGIR Conference on Research and Development in Information Retrieval<br>●Annual Meeting of the Association for Computational Linguistics<br>●IEEE/CVF Conference on Computer Vision and Pattern Recognition (CVPR) | ●IEEE Access<br>●AI and Ethics<br>●Artificial Intelligence and Law<br>●Data Mining and Knowledge Discovery<br>●Journal of the American Medical Informatics Association |
| **Covariate Shift** | ●IEEE Nuclear Science Symposium and Medical Imaging Conference (NSS/MIC)<br>●IEEE/CVF Conference on Computer Vision and Pattern Recognition (CVPR)<br>●Lecture Notes in Computer Science, Computer Vision (ECCV)<br>●IEEE International Conference on Image Processing (ICIP) | ●IEEE Access<br>●Neurocomputing<br>●IEEE Transactions on Neural Networks and Learning Systems |

(*continued on next page*)

(*continued*)

| Topic | Top-5 Most Frequent | |
|---|---|---|
| | Conferences | Journals |
| | ●International Joint Conference on Neural Networks (IJCNN) | ●Electronics<br>●Applied Soft Computing |
| **Cybersecurity Applications** | ●ACM SIGSAC Conference on Computer and Communications Security<br>●IEEE/ACM International Conference on Advances in Social Networks Analysis and Mining<br>●International Conference on Computing Methodologies and Communication (ICCMC)<br>●ACM Conference on Data and Application Security and Privacy<br>●ACM SIGKDD Conference on Knowledge Discovery & Data Mining | ●IEEE Access<br>●Expert Systems with Applications<br>●IEEE Transactions on Computer-Aided Design of Integrated Circuits and Systems<br>●IEEE Transactions on Knowledge and Data Engineering<br>●Journal of Cryptographic Engineering |
| **Systematic Preference** | ●IEEE Congress on Evolutionary Computation (CEC)<br>●Genetic and Evolutionary Computation Conference Companion<br>●Advances in Neural Information Processing Systems<br>●Annual conference on Genetic and evolutionary computation<br>●Companion Publication of the Annual Conference on Genetic and Evolutionary Computation | ●IEEE Transactions on Evolutionary Computation<br>●Applied Soft Computing<br>●Bioinformatics<br>●Information Sciences<br>●Knowledge-Based Systems<br>●Neurocomputing |
| **AI Medical Applications** | ●ACM Conference on Economics and Computation<br>●IEEE International Conference on Big Data (Big Data)<br>●IEEE International Conference on Bioinformatics and Biomedicine (BIBM)<br>●Lecture Notes in Computer Science, Medical Image Computing and Computer Assisted Intervention , MICCAI<br>●ACM Conference on Bioinformatics, Computational Biology and Health Informatics | ●Journal of the American Medical Informatics Association<br>●International Journal of Medical Informatics<br>●Journal of Biomedical Informatics<br>●Computers in Biology and Medicine<br>●Artificial Intelligence in Medicine<br>●Computer Methods and Programs in Biomedicine |
| **Information Retrieval** | ●International ACM SIGIR Conference on Research and Development in Information Retrieval<br>●ACM international conference on Information and knowledge management<br>●Web Conference<br>●ACM international conference on Web search and data mining<br>●International Conference on The Theory of Information Retrieval | ●Information Processing & Management<br>●ACM Transactions on Information Systems<br>●Information Retrieval Journal<br>●Expert Systems with Applications<br>●Knowledge and Information Systems |
| **Distribution of Training Data** | ●IEEE/CVF Conference on Computer Vision and Pattern Recognition (CVPR)<br>●ACM International Conference on Information & Knowledge Management<br>●IEEE/ACM International Conference on Advances in Social Networks Analysis and Mining<br>●Empirical Methods in Natural Language Processing<br>●International Joint Conference on Neural Networks (IJCNN)<br>●Lecture Notes in Computer Science, Computer Vision (ECCV) | ●IEEE Access<br>●IEEE Transactions on Pattern Analysis and Machine Intelligence<br>●Expert Systems with Applications<br>●IEEE Transactions on Knowledge and Data Engineering<br>●Journal of Artificial Intelligence Research<br>●Knowledge-Based Systems |
| **AI Ethical Issues** | ●AAAI/ACM Conference on AI, Ethics, and Society<br>●ACM Conference on Fairness, Accountability, and Transparency<br>●ACM SIGKDD International Conference on Knowledge Discovery & Data Mining<br>●CHI Conference on Human Factors in Computing Systems<br>●ACM Technical Symposium on Computer Science Education | ●AI and Ethics<br>●AI & SOCIETY<br>●IEEE Access<br>●Journal of the American Medical Informatics Association<br>●Nature Machine Intelligence |
| **Image Segmentation** | ●IEEE International Conference on Image Processing (ICIP)<br>●IEEE International Conference on Fuzzy Systems (FUZZ-IEEE)<br>●IEEE International Conference on Bioinformatics and Biomedicine (BIBM)<br>●IEEE International Conference on Multimedia and Expo (ICME)<br>●IEEE International Symposium on Biomedical Imaging: From Nano to Macro | ●IEEE Access<br>●Pattern Recognition<br>●IEEE Transactions on Medical Imaging<br>●IET Image Processing<br>●Neurocomputing |
| **Predictive Modeling** | ●IEEE Nuclear Science Symposium and Medical Imaging Conference (NSS/MIC)<br>●International Joint Conference on Neural Networks (IJCNN)<br>●IEEE International Conference on Systems, Man, and Cybernetics (SMC)<br>●IEEE Symposium Series on Computational Intelligence (SSCI)<br>●Asian Conference on Innovation in Technology (ASIANCON)<br>●IEEE International Conference on Data Mining (ICDM) | ●IEEE Access<br>●Journal of Chemical Information and Modeling<br>●Expert Systems with Applications<br>●Neural Computing and Applications<br>●Bioinformatics |
| **Reinforcement Learning** | ●IEEE/RSJ International Conference on Intelligent Robots and Systems (IROS)<br>●IEEE International Conference on Robotics and Automation (ICRA)<br>●International Conference on Machine Learning<br>●International Joint Conference on Neural Networks (IJCNN)<br>●Neural Information Processing Systems | ●IEEE Transactions on Neural Networks and Learning Systems<br>●Journal of Artificial Intelligence Research<br>●IEEE Transactions on Cybernetics<br>●Machine Learning<br>●Neural Networks<br>●IEEE Robotics and Automation Letters |
| **NLP** | ●Annual Meeting of the Association for Computational Linguistics<br>●ACM Conference on Fairness, Accountability, and Transparency<br>●Association for Computational Linguistics<br>●AAAI/ACM Conference on AI, Ethics, and Society<br>●IEEE International Conference on Big Data (Big Data)<br>●Advances in Neural Information Processing Systems | ●IEEE Access<br>●Expert Systems with Applications<br>●Information Processing & Management<br>●Knowledge-Based Systems<br>●ACM Transactions on Software Engineering and Methodology |
| **Unsupervised Clustering** | ●International Joint Conference on Neural Networks (IJCNN)<br>●ICASSP - IEEE International Conference on Acoustics, Speech and Signal Processing (ICASSP)<br>●IEEE International Conference on Data Mining (ICDM)<br>●IEEE International Conference on Fuzzy Systems (FUZZ-IEEE)<br>●IEEE Nuclear Science Symposium and Medical Imaging Conference (NSS/MIC) | ●IEEE Access<br>●Knowledge-Based Systems<br>●IEEE Transactions on Knowledge and Data Engineering<br>●Neurocomputing<br>●Expert Systems with Applications |
| **Computer Vision** | ●IEEE/CVF Conference on Computer Vision and Pattern Recognition (CVPR)<br>●IEEE International Conference on Image Processing (ICIP)<br>●IEEE/CVF International Conference on Computer Vision (ICCV)<br>●Lecture Notes in Computer Science, Computer Vision (ECCV)<br>●ACM International Conference on Multimedia | ●IEEE Access<br>●IEEE Transactions on Image Processing<br>●International Journal of Computer Vision<br>●Multimedia Tools and Applications<br>●IEEE Transactions on Multimedia |

## Data availability

Data will be made available on request.

## References

Adnan, M. N. (2022). On reducing the bias of random forest. *Advanced Data Mining and Applications: 18th International Conference, ADMA 2022, Brisbane, QLD, Australia, November 28–30, 2022, Proceedings, Part, II*, 187–195. https://doi.org/10.1007/978-3-031-22137-8_14

Ali, O., Murray, P. A., Momin, M., Dwivedi, Y. K., & Malik, T. (2024). The effects of artificial intelligence applications in educational settings: Challenges and strategies. *Technological Forecasting and Social Change, 199*, 123076. https://doi.org/10.1016/j.techfore.2023.123076, 123076.

Altarturi, H. H. M., Saadoon, M., & Anuar, N. B. (2023). Web content topic modeling using LDA and HTML tags. *PeerJ Computer Science, 9*, Article e1459. https://doi.org/10.7717/peerj-cs.1459

Antons, D., Breidbach, C. F., Joshi, A. M., & Salge, T. O. (2023). Computational literature reviews: Method, algorithms, and roadmap. *Organizational Research Methods, 26*(1), 107–138. https://doi.org/10.1177/1094428121991230

Ao, Z., Horváth, G.b., Sheng, C.c., Song, Y.d., & Sun, Y. D. (2023). Skill requirements in job advertisements: A comparison of skill-categorization methods based on wage regressions. *Information Processing & Management, 60*(2), Article 103185. https://doi.org/10.1016/j.ipm.2022.103185

Aralikatte, R., Sridhara, G., Gantayat, N., & Mani, S. (2018). Fault in your stars: An analysis of android app reviews. *Proceedings of the ACM India joint international conference on data science and management of data* (pp. 57–66). https://doi.org/10.1145/3152494.3152500

Asatiani, A., Malo, P., Nagbøl, P. R., Penttinen, E., Rinta-Kahila, T., & Salovaara, A. (2021). Sociotechnical envelopment of artificial intelligence: An approach to organizational deployment of inscrutable artificial intelligence systems. *Journal of the Association for Information Systems, 22*(2), 325–352. https://doi.org/10.17705/1jais.00664

Avellan, T., Sharma, S., & Turunen, M. (2020). AI for all: Defining the what, why, and how of inclusive AI. *Proceedings of the 23rd international conference on academic mindtrek* (pp. 142–144). https://doi.org/10.1145/3377290.3377317

Balgi, S., & Dukkipati, A. (2022). Contradistinguisher: A Vapnik's imperative to unsupervised domain adaptation. *IEEE Transactions on Pattern Analysis and Machine Intelligence, 44*(9), 4730–4747. https://doi.org/10.1109/TPAMI.2021.3071225

Barocas, S., & Hardt, M. (2023). *Fairness and machine learning: Limitations and opportunities*. MIT Press.

Bechky, B. A. (2003). Sharing meaning across occupational communities: The transformation of understanding on a production floor. *Organization Science, 14*(3), 312–330. https://doi.org/10.1287/orsc.14.3.312.15162

Bender, E. M., Gebru, T., McMillan-Major, A., & Shmitchell, S. (2021). On the dangers of stochastic parrots: Can language models be too big?. *Proceedings of the 2021 ACM conference on fairness, accountability, and transparency* (pp. 610–623). https://doi.org/10.1145/3442188.3445922

Bhattacharyya, S., Jha, S., Tharakunnel, K., & Westland, J. C. (2011). Data mining for credit card fraud: A comparative study. *Decision Support Systems, 50*(3), 602–613. https://doi.org/10.1016/j.dss.2010.08.008

Binns, R. (2018). Fairness in machine learning: Lessons from political philosophy. In *Conference on fairness, accountability and transparency* (pp. 149–159). https://proceedings.mlr.press/v81/binns18a.html.

Blei, D. M., Ng, A. Y., & Jordan, M. I. (2003). Latent dirichlet allocation. *Journal of machine Learning research, 3*(Jan), 993–1022.

Bolukbasi, T., Chang, K. W., Zou, J. Y., Saligrama, V., & Kalai, A. T. (2016). Man is to computer programmer as woman is to homemaker? Debiasing word embeddings. *Advances in Neural Information Processing Systems, 29*.

Borji, A., & Tanner, J. (2016). Reconciling saliency and object center-bias hypotheses in explaining free-viewing fixations. *IEEE Transactions on Neural Networks and Learning Systems, 27*(6), 1214–1226. https://doi.org/10.1016/10.1109/TNNLS.2015.2480683

Braun, V., & Clarke, V. (2006). Using thematic analysis in psychology. *Qualitative Research in Psychology, 3*(2), 77–101. https://doi.org/10.1191/1478088706qp063oa

Bruno, A., Gugliuzza, F., Pirrone, R., & Ardizzone, E. (2020). A multi-scale colour and keypoint density-based approach for visual saliency detection. *IEEE Access, 8*, 121330–121343. https://doi.org/10.1109/ACCESS.2020.3006700

Buhmann, A., & Fieseler, C. (2023). Deep learning meets deep democracy: Deliberative governance and responsible innovation in artificial intelligence. *Business Ethics Quarterly, 33*(1), 146–179. https://doi.org/10.1017/beq.2021.42

Cao, Q., Cheng, X., & Liao, S. (2022). A comparison study of topic modeling based literature analysis by using full texts and abstracts of scientific articles: A case of COVID-19 research. *Library Hi Tech, 41*(2), 543–569. https://doi.org/10.1108/LHT-03-2022-0144

Cave, S. (2020). The problem with intelligence. *Proceedings of the AAAI/ACM conference on AI, ethics, and society. AIES '20: AAAI/ACM conference on AI*. https://doi.org/10.1145/3375627.3375813. Ethics, and Society, New York NY USA.

Chawla, N. V., Lazarevic, A., Hall, L. O., & Bowyer, K. W. (2003). SMOTEBoost: Improving prediction of the minority class in boosting. *Knowledge discovery in databases: Pkdd 2003* (pp. 107–119). https://doi.org/10.1007/978-3-540-39804-2_12

Chen, Y.-J., Jiang, W.-C., Ju, M.-Y., & Hwang, K.-S. (2020). Policy sharing using aggregation trees for Q -learning in a continuous state and action spaces. *IEEE Transactions on Cognitive and Developmental Systems, 12*(3), 474–485. https://doi.org/10.1109/TCDS.2019.2926477

Chen, Y., Mahoney, C., Grasso, I., Wali, E., Matthews, A., Middleton, T., Njie, M., & Matthews, J. (2021). Gender bias and under-representation in natural language processing across human languages. *Proceedings of the 2021 AAAI/ACM conference on AI, ethics, and society* (pp. 24–34). https://doi.org/10.1145/3461702.3462530

Chenu, A., Perrin-Gilbert, N., & Sigaud, O. (2022). *Divide & conquer imitation learning* (pp. 8630–8637). IEEE/RSJ International Conference on Intelligent Robots and Systems (IROS). https://doi.org/10.1109/IROS47612.2022.9982020, 2022.

Cicek, Z. I E., & Ozturk, Z. K. (2021). Optimizing the artificial neural network parameters using a biased random key genetic algorithm for time series forecasting. *Applied Soft Computing, 102*, 107091. https://doi.org/10.1016/j.asoc.2021.107091.

Clevert, D.-A., Unterthiner, T., & Hochreiter, S. (2016). Fast and accurate deep network learning by exponential linear units (ELUs). *International Conference on Learning Representations (ICLR), San Juan, Puerto Rico (2016), pp. 1-14*.

Collins, C., Dennehy, D., Conboy, K., & Mikalef, P. (2021). Artificial intelligence in information systems research: A systematic literature review and research agenda. *International Journal of Information Management, 60*, 102383. https://doi.org/10.1016/j.ijinfomgt.2021.102383, 102383.

Crandall, J. W., & Goodrich, M. A. (2011). Learning to compete, coordinate, and cooperate in repeated games using reinforcement learning. *Machine Learning, 82*(3), 281–314. https://doi.org/10.1007/s10994-010-5192-9

Deveaud, R., SanJuan, E., & Bellot, P. (2014). Accurate and effective latent concept modeling for ad hoc information retrieval. *Document Numérique, 17*(1), 61–84. https://shs.cairn.info/article/DN_171_0061.

Dieng, A. B., Ruiz, F. J. R., & Blei, D. M. (2020). Topic modeling in embedding spaces. *Transactions of the Association for Computational Linguistics, 8*, 439–453. https://doi.org/10.1162/tacl_a_00325

Dietterich, T. G., & Kong, E. B. (1995). Machine learning bias, statistical bias, and statistical variance of decision tree algorithms. *Cités*. https://citeseerx.ist.psu.edu/document?repid=rep1&type=pdf&doi=893b204890394d1bf4f3332b4b902bfdb30a9a13.

Duan, Y., Chang, H., Huang, W., & Zhou, J. (2014). Simultaneous bias correction and image segmentation via L0 regularized mumford-shah model. *2014 IEEE international conference on image processing (ICIP)* (pp. 6–40). https://doi.org/10.1109/ICIP.2014.7025000

Dwork, C., Hardt, M., Pitassi, T., Reingold, O., & Zemel, R. (2012). Fairness through awareness. In *Proceedings of the 3rd innovations in theoretical computer science conference* (pp. 214–226). https://doi.org/10.1145/2090236.2090255

Egger, R., & Yu, J. (2022). A topic modeling comparison between LDA, NMF, Top2Vec, and BERTopic to demystify Twitter posts. *Frontiers in Sociology, 7*, Article 886498. https://doi.org/10.3389/fsoc.2022.886498

Eitel-Porter, R. (2021). Beyond the promise: Implementing ethical AI. *AI and Ethics, 1*(1), 73–80. https://doi.org/10.1007/s43681-020-00011-6

ElShawi, R., Sherif, Y., Al-Mallah, M., & Sakr, S. (2021). Interpretability in healthcare: A comparative study of local machine learning interpretability techniques. *Computational Intelligence, 37*(4), 1633–1650. https://doi.org/10.1111/coin.12410

Erhan, D., Courville, A., Bengio, Y., & Vincent, P. (2010). Why does unsupervised pre-training help deep learning?. *Proceedings of the thirteenth international conference on artificial intelligence and statistics* (pp. 201–208). https://proceedings.mlr.press/v9/erhan10a.html.

Erzurum Cicek, Z. I., & Kamisli Ozturk, Z. (2021). Optimizing the artificial neural network parameters using a biased random key genetic algorithm for time series forecasting. *Applied Soft Computing, 102*, Article 107091. https://doi.org/10.1016/j.asoc.2021.107091

Fang, Y., Lin, W., Fang, Z., Lei, J., Le Callet, P., & Yuan, F. (2014). Learning visual saliency for stereoscopic images. *2014 IEEE international conference on multimedia and expo workshops (ICMEW)* (pp. 1–6). https://doi.org/10.1109/ICMEW.2014.6890709

Fast, E., Vachovsky, T., & Bernstein, M. (2016). Shirtless and dangerous: Quantifying linguistic signals of gender bias in an online fiction writing community. *Proceedings of the International AAAI Conference on Web and Social Media, 10*(1), 112–120. https://doi.org/10.1609/icwsm.v10i1.14744

Feffer, M., Sinha, A., Deng, W. H., Lipton, Z. C., & Heidari, H. (2024). Red-teaming for generative AI: Silver bullet or security theater? In *arXiv [cs.CY]. arXiv*. http://arxiv.org/abs/2401.15897.

Felzmann, H., Fosch-Villaronga, E., Lutz, C., & Tamò-Larrieux, A. (2020). Towards transparency by design for artificial intelligence. *Science and Engineering Ethics, 26*(6), 3333–3361. https://doi.org/10.1007/s11948-020-00276-4

Ferrara, E. (2023). Fairness and bias in artificial intelligence: A brief survey of sources, impacts, and mitigation strategies. *Science, 6*(1), 3. https://doi.org/10.3390/sci6010003

Fiore, U., De Santis, A., Perla, F., Zanetti, P., & Palmieri, F. (2019). Using generative adversarial networks for improving classification effectiveness in credit card fraud detection. *Information Sciences, 479*, 448–455. https://doi.org/10.1016/j.ins.2017.12.030.

Forsyth, S., Dalton, B., Foster, E. H., Walsh, B., Smilack, J., & Yeh, T. (2021). Imagine a more ethical AI: Using stories to develop teens' awareness and understanding of artificial intelligence and its societal impacts. *2021 conference on research in equitable* *and sustained participation in engineering, computing, and technology (RESPECT)* (pp. 1–2). https://doi.org/10.1109/RESPECT51740.2021.9620549

Fosch-Villaronga, E., Poulsen, A., Søraa, R. A., & Custers, B. H. M. (2021). A little bird told me your gender: Gender inferences in social media. *Information Processing & Management, 58*(3), Article 102541. https://doi.org/10.1016/j.ipm.2021.102541

Freyne, J., Coyle, L., Smyth, B., & Cunningham, P. (2010). Relative status of journal and conference publications in computer science. *Communications of the ACM, 53*(11), 124–132. https://doi.org/10.1145/1839676.1839701

Friedman, B., & Nissenbaum, H. (1996). Bias in computer systems. *ACM Transactions on Information and System Security, 14*(3), 330–347. https://doi.org/10.1145/230538.230561

Fu, H., Liu, J., Wu, G., Xu, Y., & Sutcliffe, G. (2022). Improving probability selection based weights for satisfiability problems. *Knowledge-Based Systems, 245*, Article 108572. https://doi.org/10.1016/j.knosys.2022.108572

Gharghabi, S., Imani, S., Bagnall, A., Darvishzadeh, A., & Keogh, E. (2018). Matrix Profile XII: MPdist: A novel time series distance measure to allow data mining in more challenging scenarios. *2018 IEEE international conference on data mining (ICDM)* (pp. 965–970). https://doi.org/10.1109/ICDM.2018.00119

Gmyrek, P., Lutz, C., & Newlands, G. (2025). A technological construction of society: Comparing GPT-4 and human respondents for occupational evaluation in the UK. *British Journal of Industrial Relations, 63*(1), 180–208. https://doi.org/10.1111/bjir.12840

Grasas, A., Juan, A. A., Faulin, J., De Armas, J., & Ramalhinho, H. (2017). Biased randomization of heuristics using skewed probability distributions: A survey and some applications. *Computers & Industrial Engineering, 110*, 216–228. https://doi.org/10.1016/j.cie.2017.06.019

Griffin, T. A., Green, B. P., & Welie, J. V. M. (2023). The ethical agency of AI developers. *AI and Ethics*. https://doi.org/10.1007/s43681-022-00256-3

Guo, H., & Viktor, H. L. (2004). Learning from imbalanced data sets with boosting and data generation: The DataBoost-IM approach. *SIGKDD Explor, 6*(1), 30–39. https://doi.org/10.1145/1007730.100773. Newsl.

Gupta, R., & Alam, T. (2023). A deep neural network with hybrid spotted hyena optimizer and grasshopper optimization algorithm for copy move forgery detection. *Multimedia Tools and Applications, 82*(16), 24547–24572. https://doi.org/10.1007/s11042-022-14163-6

Hagen, L. (2018). Content analysis of e-petitions with topic modeling: How to train and evaluate LDA models? *Information Processing & Management, 54*(6), 1292–1307. https://doi.org/10.1016/j.ipm.2018.05.006

Hall, P., & Ellis, D. (2023). A systematic review of socio-technical gender bias in AI algorithms. *Online Information Review, 47*(7), 1264–1279. https://doi.org/10.1108/OIR-08-2021-0452

Hannigan, T. R., Haans, R. F., Vakili, K., Tchalian, H., Glaser, V. L., Wang, M. S., … Jennings, P. D. (2019). Topic modeling in management research: Rendering new theory from textual data. *Academy of Management Annals, 13*(2), 586–632. https://doi.org/10.5465/annals.2017.0099.

Higgins, J. P. T., Altman, D. G., Gøtzsche, P. C., Jüni, P., Moher, D., Oxman, A. D., Savovic, J., Schulz, K. F., Weeks, L., & Sterne, J. A. C. (2011). The cochrane Collaboration's tool for assessing risk of bias in randomised trials. *BMJ, 343*, Article d5928. https://doi.org/10.1136/bmj.d5928

Hu, Q., He, Q., Huang, H., Chiew, K., & Liu, Z. (2016). A formalized framework for incorporating expert labels in crowdsourcing environment. *Journal of Intelligent Information Systems, 47*(3), 403–425. https://doi.org/10.1007/s10844-015-0371-6

Hu, Qian, & Rangwala, H. Metric-free individual fairness with cooperative contextual bandits. https://doi.org/10.1109/ICDM50108.2020.00027.

Islam, S., & Hassanzadeh Amin, S. (2020). Prediction of probable backorder scenarios in the supply chain using distributed random forest and gradient boosting machine learning techniques. *Journal of Big Data, 7*(1), s40537. https://doi.org/10.1186/s40537-020-00345-2, 020.

Jarrahi, M. H., Lutz, C., Boyd, K., Østerlund, C., & Willis, M. (2023a). AI in the work context. *Journal of the Association for Information Science and Technology, 74*(3). https://doi.org/10.1002/asi.24730

Jarrahi, M. H., Memariani, A., & Guha, S. (2023b). The principles of data-centric AI. *Communications of the ACM, 66*(8), 84–92. https://doi.org/10.1145/3571724

Ji, Z., Liu, J., & Li, G. (2014). A fuzzy clustering algorithm via enhanced spatially constraint for brain MR image segmentation. *2014 international conference on Orange technologies* (pp. 105–108). https://doi.org/10.1109/ICOT.2014.6956610

Jiang, L., Xu, M., Liu, T., Qiao, M., & Wang, Z. (2018). DeepVS: A deep learning based video saliency prediction approach. *Computer Vision – ECCV*, 625–642, 2018 https://openaccess.thecvf.com/content_ECCV_2018/html/Lai_Jiang_DeepVS_A_Deep_ECCV_2018_paper.html.

Jin, L., Long, X., Zhang, K., Lin, Y. R., & Joshi, J. (2016). Characterizing users' check-in activities using their scores in a location-based social network. *Multimedia Systems, 22*(1), 87–98. https://doi.org/10.1007/s00530-014-0395-8

Juntu, J., Sijbers, J., Van Dyck, D., & Gielen, J. (2005). Bias field correction for MRI images. *Computer Recognition Systems*, 543–551. https://doi.org/10.1007/3-540-32390-2_64

Karami, A. (2015). In A. Gangopadhyay (Ed.), *Fuzzy topic modeling for medical corpora*. Baltimore County]: University of Maryland. http://libproxy.lib.unc.edu/login?url=https://www.proquest.com/dissertations-theses/fuzzy-topic-modeling-medical-corpora/docview/1721469756/se-2.

Karami, A., Lundy, M., Webb, F., & Dwivedi, Y. K. (2020). Twitter and research: A systematic literature review through text mining. *IEEE Access, 8*, 67698–67717. https://doi.org/10.1109/ACCESS.2020.2983656

Karami, A., Qiao, Z., Zhang, X., Kharrazi, H., Bozorgi, P., & Bozorgi, A. (2024). Health use cases of AI chatbots: Identification and analysis of ChatGPT prompts in social media discourses. *Big Data and Cognitive Computing, 8*(10), 130. https://doi.org/10.3390/bdcc8100130

Karami, A., White, C. N., Ford, K., Swan, S., & Yildiz Spinel, M. (2020). Unwanted advances in higher education:uncovering sexual harassment experiences in academia with text mining. *Information Processing & Management, 57*(2), Article 102167. https://doi.org/10.1016/j.ipm.2019.102167

Kärkkäinen, K., & Joo, J. (2021). FairFace: Face attribute dataset for balanced race, gender, and age for bias measurement and mitigation. *2021 IEEE winter conference on applications of computer vision (WACV)* (pp. 1547–1557). https://openaccess.thecvf.com/content/WACV2021/html/Karkkainen_FairFace_Face_Attribute_Dataset_for_Balanced_Race_Gender_and_Age_WACV_2021_paper.html.

Kordzadeh, N., & Ghasemaghaei, M. (2022). Algorithmic bias: Review, synthesis, and future research directions. *European Journal of Information Systems, 31*(3), 388–409. https://doi.org/10.1080/0960085X.2021.1927212

Kumar, S., Biswas, S. K., & Devi, D. (2019). TLUSBoost algorithm: A boosting solution for class imbalance problem. *Soft Computing, 23*(21), 10755–10767. https://doi.org/10.1007/s00500-018-3629-4

Lam-Adesina, A. M., & Jones, G. J. F. (2001). Applying summarization techniques for term selection in relevance feedback. *Proceedings of the 24th annual international ACM SIGIR conference on research and development in information retrieval* (pp. 1–9). https://doi.org/10.1145/383952.383953

Levac, D., Colquhoun, H., & O'Brien, K. K. (2010). Scoping studies: Advancing the methodology. *Implementation Science, 5*, 1–9. https://doi.org/10.1186/1748-5908-5-69

Li, X., Fang, M., Li, H., & Wu, J. (2020). Learning domain invariant unseen features for generalized zero-shot classification. *Knowledge-Based Systems, 206*, Article 106378. https://doi.org/10.1016/j.knosys.2020.106378

Li, E., Feng, H., Zhou, H., Li, X., Zhai, Y., Zhang, S., & Fu, Y. (2019). Model learning for two-wheeled robot self-balance control. In *IEEE international conference on robotics and biomimetics (ROBIO)* (pp. 1582–1587). https://doi.org/10.1109/ROBIO49542.2019.8961382, 2019.

Li, C., Huang, R., Ding, Z., Gatenby, J. C., Metaxas, D. N., & Gore, J. C. (2011). A level set method for image segmentation in the presence of intensity inhomogeneities with application to MRI. *IEEE Transactions on Image Processing, 20*(7), 2007–2016. https://doi.org/10.1109/TIP.2011.2146190

Limthong, K., Fukuda, K., Ji, Y., & Yamada, S. (2015). Weighting technique on multi-timeline for machine learning-based anomaly detection system. In *2015 international conference on computing, communication and security (ICCCS)* (pp. 1–6). https://doi.org/10.1109/CCCS.2015.7374168

Liu, X., Zhou, Y., & Wang, Z. (2022). Preference access of users' cancer risk perception using disease-specific online medical inquiry texts. *Information Processing & Management, 59*(1), Article 102737. https://doi.org/10.1016/j.ipm.2021.102737

Lu, Y., Mei, Q., & Zhai, C. (2011). Investigating task performance of probabilistic topic models: An empirical study of PLSA and LDA. *Information Retrieval, 14*(2), 178–203. https://doi.org/10.1007/s10791-010-9141-9

Lutz, C. (2024). Social inequalities and artificial intelligence: How digital inequality scholarship enhances our understanding. In D. Brzeziński, K. Filipek, K. Piwowar, & M. Winiarska-Brodowska (Eds.), *Algorithms, artificial intelligence and beyond* (pp. 193–210). Routledge.

Maclure, J. (2021). AI, explainability and public reason: The argument from the limitations of the human mind. *Minds and Machines, 31*(3), 421–438. https://doi.org/10.1007/s11023-021-09570-x

Madaio, M. A., Stark, L., Wortman Vaughan, J., & Wallach, H. (2020). Co-designing checklists to understand organizational challenges and opportunities around fairness in AI. In *Proceedings of the 2020 CHI conference on human factors in computing systems* (pp. 1–14). https://doi.org/10.1145/3313831.3376445

Malek, M. A. (2022). Criminal courts' artificial intelligence: The way it reinforces bias and discrimination. *AI and Ethics, 2*(1), 233–245. https://doi.org/10.1007/s43681-022-00137-9

McCallum, A. K. (2002). Mallet: A machine learning for languagetoolkit. https://cir.nii.ac.jp/crid/1573105974103526144.

Mehrabi, N., Morstatter, F., Saxena, N., Lerman, K., & Galstyan, A. (2022). A survey on bias and fairness in machine learning. *ACM Computing Surveys, 45*(6), 1–35. https://doi.org/10.1145/3457607

Mittelstadt, B. D., Allo, P., Taddeo, M., Wachter, S., & Floridi, L. (2016). The ethics of algorithms: Mapping the debate. *Big Data & Society, 3*(2), 1–21. https://doi.org/10.1177/2053951716679679

Miyata, Y., Ishita, E., Yang, F., Yamamoto, M., Iwase, A., & Kurata, K. (2020). Knowledge structure transition in library and information science: Topic modeling and visualization. *Scientometrics, 125*(1), 665–687. https://doi.org/10.1007/s11192-020-03657-5

Moghanian, S., Saravi, F. B., Javidi, G., & Sheybani, E. O. (2020). GOAMLP: Network intrusion detection with multilayer perceptron and grasshopper optimization algorithm. *IEEE Access, 8*, 215202–215213. https://doi.org/10.1109/ACCESS.2020.3040740

Mohammadi, M., Al-Fuqaha, A., Sorour, S., & Guizani, M. (2018). Deep learning for IoT big data and streaming analytics: A survey. *IEEE Communications Surveys & Tutorials, 20*(4), 2923–2960. https://doi.org/10.1109/COMST.2018.2844341

Ncir, N., Sebbane, S., & El Akchioui, N. (2022). A novel intelligent technique based on metaheuristic algorithms and artificial neural networks: Application on a photovoltaic panel. *2022 2nd international conference on innovative research in applied science, engineering and technology (IRASET)* (pp. 1–8). https://doi.org/10.1109/IRASET52964.2022.9738106

Ntoutsi, E., Fafalios, P., Gadiraju, U., Iosifidis, V., Nejdl, W., Vidal, M.-E., Ruggieri, S., Turini, F., Papadopoulos, S., Krasanakis, E., & Others. (2020). Bias in data-driven artificial intelligence systems—An introductory survey. *Wiley Interdisciplinary*

*Reviews. Data Mining and Knowledge Discovery, 10*(3), Article e1356. https://doi.org/10.1002/widm.1356

Olteanu, A., Castillo, C., Diaz, F., & Kıcıman, E. (2019). Social data: Biases, methodological pitfalls, and ethical boundaries. *Frontiers in Big Data, 2*, 13. https://doi.org/10.3389/fdata.2019.00013

Ouadrhiri, A. E., & Abdelhadi, A. (2021). Differential privacy for fair deep learning models. *2021 IEEE international systems conference (SysCon)* (pp. 1–6). https://doi.org/10.1109/SysCon48628.2021.9591252

Pagano, T. P., Loureiro, R. B., Lisboa, F. V. N., Peixoto, R. M., Guimarães, G. A. S., Cruz, G. O. R., Araujo, M. M., Santos, L. L., Cruz, M. A. S., & Oliveira, E. L. S. (2023). Machine learning models: A systematic review on datasets, tools, fairness metrics, and identification and mitigation methods. *Big Data and Cognitive Computing, 7*(1), 15. https://doi.org/10.3390/bdcc7010015

Page, M. J., Moher, D., Bossuyt, P. M., Boutron, I., Hoffmann, T. C., Mulrow, & McKenzie, J. E. (2021). PRISMA 2020 explanation and elaboration: Updated guidance and exemplars for reporting systematic reviews. *BMJ, 372*, Article n160. https://doi.org/10.1136/bmj.n160

Papamartzivanos, D., Gómez Mármol, F., & Kambourakis, G. (2018). Dendron : Genetic trees driven rule induction for network intrusion detection systems. *Future Generation Computer Systems: FGCS, 79*, 558–574. https://doi.org/10.1016/j.future.2017.09.056

Raghuwanshi, B. S., & Shukla, S. (2020). SMOTE based class-specific extreme learning machine for imbalanced learning. *Knowledge-Based Systems, 187*, Article 104814. https://doi.org/10.1016/j.knosys.2019.06.022

Raji, I. D., Smart, A., White, R. N., Mitchell, M., Gebru, T., Hutchinson, B., … Barnes, P. (2020). Closing the AI accountability gap: Defining an end-to-end framework for internal algorithmic auditing. In *Proceedings of the 2020 conference on fairness, accountability, and transparency* (pp. 33–44). https://doi.org/10.1145/3351095.3372873

Sakai, Y. (2020). Quantizaiton for deep neural network training with 8-bit dynamic fixed point. *2020 7th international conference on soft computing & machine intelligence (ISCMI)* (pp. 126–130). https://doi.org/10.1109/ISCMI51676.2020.9311563

Sandvig, C., Hamilton, K., Karahalios, K., & Langbort, C. (2014). Auditing algorithms: Research methods for detecting discrimination on internet platforms. *Data and Discrimination: Converting Critical Concerns into Productive Inquiry, 22*, 4349–4357, 2014 https://www.kevinhamilton.org/share/papers/Auditing%20Algorithms%20–%20Sandvig%20–%20ICA%202014%20Data%20and%20Discrimination%20Preconference.pdf.

Sarker, S., University of, V., Chatterjee, S., Xiao, X., & Elbanna, A. (2019). The sociotechnical axis of cohesion for the IS discipline: Its historical legacy and its continued relevance. *MIS Quarterly, 43*(3), 695–719.

Sawyer, S., & Jarrahi, M. H. (2014). Sociotechnical approaches to the study of information systems. In *Computing handbook, third edition: Information systems and information technology* (p. 5). CRC Press, 1.

Scholer, F., Shokouhi, M., Billerbeck, B., & Turpin, A. (2008). Using clicks as implicit judgments: Expectations versus observations. *Proceedings of the IR research, 30th European conference on advances in information retrieval* (pp. 28–39). https://doi.org/10.1007/978-3-540-78646-7_6

Schwartz, R., Vassilev, A., Greene, K., Perine, L., Burt, A., & Hall, P. (2022). *Towards a standard for identifying and managing bias in artificial intelligence*. U.S.: National Institute of Standards and Technology. https://doi.org/10.6028/nist.sp.1270

Shao, Y., Mohanty, A. F., Ahmed, A., Weir, C. R., Bray, B. E., Shah, R. U., Redd, D., & Zeng-Treitler, Q. (2016). Identification and use of frailty indicators from text to examine associations with clinical outcomes among patients with heart failure. In *AMIA ... annual symposium proceedings/AMIA symposium. AMIA symposium* (pp. 1110–1118), 2016 https://pmc.ncbi.nlm.nih.gov/articles/PMC5333331/.

Shneiderman, B. (2022). *Human-centered AI*. Oxford University Press.

Silva, K. D., Lee, W. K., Forbes, A., Demmer, R. T., Barton, C., & Enticott, J. (2020). Use and performance of machine learning models for type 2 diabetes prediction in community settings: A systematic review and meta-analysis. *International Journal of Medical Informatics, 143*, Article 104268. https://doi.org/10.1016/j.ijmedinf.2020.104268

Simon, V., Rabin, N., & Gal, H. C.-B. (2023). Utilizing data driven methods to identify gender bias in LinkedIn profiles. *Information Processing & Management, 60*(5), Article 103423. https://doi.org/10.1016/j.ipm.2023.103423

Snyder, H. (2019). Literature review as a research methodology: An overview and guidelines. *Journal of Business Research, 104*, 333–339. https://doi.org/10.1016/j.jbusres.2019.07.039

Solomonides, A. E., Koski, E., Atabaki, S. M., Weinberg, S., McGreevey, J. D., Kannry, J. L., Petersen, C., & Lehmann, C. U. (2022). Defining AMIA's artificial intelligence principles. *Journal of the American Medical Informatics Association: JAMIA, 29*(4), 585–591. https://doi.org/10.1093/jamia/ocac006

Soprano, M., Roitero, K., La Barbera, D., Ceolin, D., Spina, D., Demartini, G., & Mizzaro, S. (2024). Cognitive biases in fact-checking and their countermeasures: A review. *Information Processing & Management, 61*(3), Article 103672. https://doi.org/10.1016/j.ipm.2024.103672

Suresh, H., & Guttag, J. (2021). A framework for understanding sources of harm throughout the machine learning life cycle. *Equity and Access in Algorithms, Mechanisms, and Optimization*, 1–9. https://doi.org/10.1145/3465416.3483305

Suri, J. S., Bhagawati, M., Paul, S., Protogeron, A., Sfikakis, P. P., Kitas, G. D., Khanna, N. N., Ruzsa, Z., Sharma, A. M., Saxena, S., Faa, G., Paraskevas, K. I., Laird, J. R., Johri, A. M., Saba, L., & Kalra, M. (2022). Understanding the bias in machine learning systems for cardiovascular disease risk assessment: The first of its kind review. *Computers in Biology and Medicine, 142*, Article 105204. https://doi.org/10.1016/j.compbiomed.2021.105204

Swazinna, P., Udluft, S., & Runkler, T. (2021). Overcoming model bias for robust offline deep reinforcement learning. *Engineering Applications of Artificial Intelligence, 104*, 104366. https://doi.org/10.1016/j.engappai.2021.104366.

Syed, S., & Spruit, M. (2017). Full-text or abstract? Examining topic coherence scores using latent dirichlet allocation. In *2017 IEEE international conference on data science and advanced analytics (DSAA)*. https://doi.org/10.1109/dsaa.2017.61. Tokyo, Japan.

Tachtler, F., Aal, K., Ertl, T., Diethei, D., Niess, J., Khwaja, M., Talhouk, R., Vilaza, G. N., Lazem, S., Singh, A., Barry, M., Wulf, V., & Fitzpatrick, G. (2021). Artificially intelligent technology for the margins: A multidisciplinary design agenda. *Extended Abstracts of the 2021 CHI Conference on Human Factors in Computing Systems*, 1–7. https://doi.org/10.1145/3411763.3441333

Tao, X., Zheng, Y., Chen, W., Zhang, X., Qi, L., Fan, Z., & Huang, S. (2022). SVDD-based weighted oversampling technique for imbalanced and overlapped dataset learning. *Information Sciences, 588*, 13–51. https://doi.org/10.1016/j.ins.2021.12.066

Tranfield, D., Denyer, D., & Smart, P. (2003). Towards a methodology for developing evidence-informed management knowledge by means of systematic review. *British Journal of Management, 14*(3), 207–222. https://doi.org/10.1111/1467-8551.00375

Tseng, P.-H., Carmi, R., Cameron, I. G. M., Munoz, D. P., & Itti, L. (2009). Quantifying center bias of observers in free viewing of dynamic natural scenes. *Journal of Vision, 9*(7), 4. https://doi.org/10.1167/9.7.4

Van Altena, A. J., Moerland, P. D., Zwinderman, A. H., & Olabarriaga, S. D. (2016). Understanding big data themes from scientific biomedical literature through topic modeling. *Journal of Big Data, 3*(1). https://doi.org/10.1186/s40537-016-0057-0

Vayansky, I., & Kumar, S. A. P. (2020). A review of topic modeling methods. *Information Systems, 94*, Article 101582. https://doi.org/10.1016/j.is.2020.101582

Waltman, L., & van Eck, N. J. (2012). A new methodology for constructing a publication-level classification system of science. *Journal of the American Society for Information Science and Technology, 63*(12), 2378–2392. https://doi.org/10.1002/asi.22748

Wang, Z., Du, B., Zhang, L., Zhang, L., & Jia, X. (2017). A novel semisupervised active-learning algorithm for hyperspectral image classification. *IEEE Transactions on Geoscience and Remote Sensing: A Publication of the IEEE Geoscience and Remote Sensing Society, 55*(6), 3071–3083. https://doi.org/10.1109/TGRS.2017.2650938

Webster, J., & Watson, R. T. (2022). Analyzing the past to prepare for the future: Writing a literature review. *MIS Quarterly, 26*(2). xiii-xxiii.

Williams, B. A., Brooks, C. F., & Shmargad, Y. (2018). How algorithms discriminate based on data they lack: Challenges, solutions, and policy implications. *Journal of Information Policy, 8*, 78–115. https://doi.org/10.5325/jinfopoli.8.2018.0078

Wolf, M. J., Miller, K., & Grodzinsky, F. S. (2017). Why we should have seen that coming. *ACM SIGCAS Computers and Society, 47*(3), 54–64. https://doi.org/10.1145/3144592.3144598

Wu, M., Li, Q., Zhang, J., & Hou, J. (2022). Label aggregation with clustering for biased crowdsourced labeling. *2022 14th international conference on machine learning and computing (ICMLC)* (pp. 165–169). https://doi.org/10.1145/3529836.3529861

Wu, T., Yao, M., & Yang, J. (2017). Dolphin swarm extreme learning machine. *Cognitive Computation, 9*(2), 275–284. https://doi.org/10.1007/s12559-017-9451-y

Xu, Y., Yang, Y., Han, J., Wang, E., Zhuang, F., Yang, J., & Xiong, H. (2019). NeuO: Exploiting the sentimental bias between ratings and reviews with neural networks. *Neural Networks: The Official Journal of the International Neural Network Society, 111*, 77–88. https://doi.org/10.1016/j.neunet.2018.12.011

Xue, J., Wang, Y.-C., Wei, C., Liu, X., Woo, J., & Kuo, C.-C. J. (2023). Bias and fairness in chatbots: An overview. *arXiv [cs.CL]. arXiv*. http://arxiv.org/abs/2309.08836.

Yang, Y., Zhang, X., Yang, M., & Deng, C. (2023). Adaptive bias-aware feature generation for generalized zero-shot learning. *IEEE Transactions on Multimedia, 25*, 280–290. https://doi.org/10.1109/TMM.2021.3125134

Zhang, F., Bai, L., & Gao, F. (2009). A user trust-based collaborative filtering recommendation algorithm. *Information and Communications Security*, 411–424. https://doi.org/10.1007/978-3-642-11145-7_32

Zhang, H., Chu, X., Asudeh, A., & Navathe, S. B. (2021). OmniFair: A declarative system for model-agnostic group fairness in machine learning. *Proceedings of the 2021 international conference on management of data* (pp. 2076–2088). https://doi.org/10.1145/3448016.3452787

Zhang, D., Luo, T., & Wang, D. (2016). Learning from LDA using deep neural networks. In *Natural language understanding and intelligent applications* (pp. 657–664). Springer. https://doi.org/10.1007/978-3-319-50496-4_59.

Zhao, H., Liu, F., Li, L., & Luo, C. (2018). A novel softplus linear unit for deep convolutional neural networks. *Applied Intelligence, 48*(7), 1707–1720. https://doi.org/10.1007/s10489-017-1028-7

Zhao, Y., Xu, T., Liu, X., Guo, D., Hu, Z., Liu, H., & Li, Y. (2022). Visual feature synthesis with semantic reconstructor for traditional and generalized zero-shot object classification. *International Journal of Intelligent Systems, 37*(5), 2934–2951. https://doi.org/10.1002/int.22811